\documentclass[range]{ar2e}
\input psfig.sty

\usepackage{amsmath,amssymb}
\usepackage{graphicx}
\usepackage{natbib}
\usepackage{setspace}

\usepackage[wide]{sidecap}

\newcommand {\chisq}    {\mbox{$\chi^2$}}

\newcommand {\Rout} {{\mbox{$R_{\rm out}$}}}
\newcommand {\Rside} {{\mbox{$R_{\rm side}$}}}
\newcommand {\Rlong} {{\mbox{$R_{\rm long}$}}}

\newcommand {\RoSq} {{\mbox{$R_{\rm out}^2$}}}
\newcommand {\RsSq} {{\mbox{$R_{\rm side}^2$}}}

\newcommand {\RosSq} {{\mbox{$R_{\rm out,side}^2$}}}

\newcommand {\Npart}   {{\mbox{$N_{\rm part}$}}}
\newcommand {\Npthree} {{\mbox{$N_{\rm part}^{1/3}$}}}

\newcommand {\dNthree} {{\mbox{$(dN_{\rm ch}/d\eta)^{1/3}$}}}
\newcommand {\YYK}     {{\mbox{${Y_{\rm YK}}$}}}
\newcommand {\roots}   {{\mbox{$\sqrt{s_{\rm NN}}$}}}
\newcommand {\xp}      {{\mbox{${\bf xp}$}}}
\newcommand {\phipair} {{\mbox{$\phi_{\rm pair}$}}}

\begin{document}


\title{Femtoscopy in Relativistic Heavy Ion Collisions: Two Decades of Progress}

\author{
Michael Annan Lisa \affiliation{Department of Physics, 
The Ohio State University, Columbus, Ohio, 43210; email lisa@mps.ohio-state.edu}
Scott Pratt \affiliation{Department of Physics and Astronomy, 
Michigan State University, East Lansing, Michigan, 48824; email pratts@pa.msu.edu}
Ron Soltz\affiliation{N-Division, Livermore National Laboratory, 
7000 East Avenue, Livermore, California, 94550; email soltz1@llnl.gov}
Urs Wiedemann\affiliation{Theory Division, CERN, Geneva, Switzerland; email urs.wiedemann@cern.ch}}

\markboth{Lisa et al.}{Femtoscopy in Heavy Ion Collisions}

\begin{keywords}
HBT, intensity interferometry, 
heavy ion Collisions, femtoscopy
\end{keywords}


\begin{singlespace}

\begin{abstract}
Analyses of two-particle correlations have provided the chief means for
determining spatio-temporal characteristics of relativistic heavy ion
collisions.  We discuss the theoretical formalism behind these studies
and the experimental methods used in carrying them out.  
Recent results from RHIC are put into context in a systematic
review of correlation measurements performed over the past two decades.
The current understanding of these results is discussed in terms of
model comparisons and overall trends.
\end{abstract}

\maketitle


\section{Introduction}
\label{sec:introduction}

The study of nucleus-nucleus collisions at ultra-relativistic energies aims to characterize the dynamical processes by which matter at extreme densities is
produced and the fundamental properties that this matter exhibits. In
nucleus-nucleus collisions, how do partonic equilibration processes proceed?
For how long, over which spatial extension, and at which density is a QCD
equilibration state approached, and what are its properties?  Particle
densities attained during a heavy ion collision are expected to exceed
significantly the inverse volume of a hadron. This implies that the high
temperature phase of QCD, the Quark Gluon Plasma, comes within experimental
reach. Chiral symmetry restoration and deconfinement phase transition are
testable in heavy ion collisions. However, the experimental study of QCD at
high temperatures and densities is complicated by the short lifetime and
mesoscopic extension of the produced system. Femtoscopy, the spatio-temporal
characterization of the collision region on the femtometer scale, is needed to
frame any discussion of dynamical equilibration processes.

The Relativistic Heavy Ion Collider (RHIC) just completed the
first part of a dedicated experimental heavy ion program.
Center of mass energies ($\sqrt{s_{NN}} = 200$ GeV) exceeded
those of previous fixed target experiments by a factor 10.
The current discussion of RHIC data focuses mainly on several
qualitatively novel phenomena that all support the picture
that dense and rapidly equilibrating QCD matter is produced in 
the collision 
region~\cite{Adams:2005dq,Adcox:2004mh,Back:2004je,Arsene:2004fa}. 
In particular, identified 
single inclusive hadron spectra appear to emerge from a 
common flow field whose size and dependence on transverse 
momentum and azimuth is consistent with expectations that 
the produced matter is a locally equilibrated, almost ideal 
fluid of very small viscosity. Moreover, high-$p_T$ hadron 
spectra show a centrality dependent, strong suppression in 
Au+Au collisions, but not in a d+Au control experiment, 
indicating that even the hardest partons produced in the
collision participate significantly in equilibration processes.
In short, experiments at RHIC have demonstrated already that 
heavy ion collisions produce dense and equilibrating matter, 
and that controlled experimentation of this matter is possible 
using a large variety of probes.

Despite these successes, numerous questions remain concerning the state of the
matter produced in these collisions. Most notably, the equation of state is far
from being determined, and issues concerning chiral symmetry restoration are
largely unresolved. Addressing these fundamental questions about bulk matter
requires a detailed understanding of the dynamics and chemistry of the collision,
which can only be acquired by thorough and coordinated analyses of data and
theory.  In particular, spatio-temporal aspects of the reaction need to be
experimentally addressed. The small size, $\sim 10^{-14}$ m, and transient
nature, $\sim 10^{-22}$ seconds, of the reactions preclude direct measurement
of times or positions. Instead, femtoscopy must exploit measurements of
asymptotic momenta. Correlations of two final-state particles at small relative
momentum provide the most direct link to the size and lifetime of subatomic
reactions~\cite{Boal:1990yh,Bauer:1993wq,Heinz:1999rw,Wiedemann:1999qn,Csorgo:1999sj,Weiner:1999th,Tomasik:2002rx,Alexander:2003ug}. 
Because correlations from either interactions or from identity interference are
stronger for smaller separations in space-time, spatio-temporal information can
be most easily extracted for small sources, unlike the limitations of
microscopes and telescopes.

The interference of two particles emitted from chaotic sources was first
applied by Hanbury-Brown and Twiss \cite{HBT1,HBT2}, where photons were
exploited to determine source sizes for both laboratory and stellar sources in
the 1950s and 1960s. Correlations of identical pions were shown to be sensitive
to source dimensions in proton-antiproton collisions by Goldhaber, Goldhaber,
Lee, and Pais in 1960 \cite{Goldhaber:1960sf}. In the 1970s, these methods were refined
by Kopylov and Podgoretsky \cite{Kopylov:1972qw,Kopylov:1974uc,Kopylov:1974a,Kopylov:1974b}, Koonin, \cite{Koonin:1977fh},
and Gyulassy \cite{Gyulassy:1979yi}, and other classes of correlations were
shown to be useful for source-size measurements, such as strong and Coulomb
interactions. Bevalac analyses showed that interferometry was truly capable of
quantitatively determining spatial and temporal source dimensions
\cite{Zajc:1984vb,Fung:1978} and providing a stringent test of dynamical models
\cite{Humanic:1985qx}. Throughout the last 25 years this phenomenology has
developed into a precision tool for heavy ion collisions. Whereas hadronic
sources are short lived and one measures correlations of the momentum of
outgoing particles, stars are long lived and require experimental filters to
enforce the approximate simultaneity of the two photons. Although
the theory for these two classes of measurement are very different
\cite{Baym:1997ce,Kopylov:1976}, the heavy-ion community often 
uses the term {\it HBT}, in reference to Hanbury-Brown and Twiss's original work with
photons, to refer to any type of analysis related to size and shape that uses two particles at small relative momentum. To some, however, the term HBT refers only to identical-particle
interferometry. Following Lednicky, we will employ the term {\it femtoscopy}
\cite{Lednicky:1990pu,Lednicky:2002fq} to denote any measurement that provides spatio-temporal
information, including coalescence analyses.

Femtoscopic measurements from truly relativistic heavy ion collisions were first reported almost twenty years ago. Since then, measurements have been performed for collisions at energies spanning two orders of magnitude. In a double sense, then, this review examines recent RHIC results within the larger context of two decades' worth of femtoscopy. The theory and phenomenology of correlation femtoscopy are reviewed in the next
section, with particular emphasis on describing the approximations used to
derive the connection between spatio-temporal aspects of the emission function
and correlations constructed from final-state momenta. Experimental methods and
techniques are correspondingly reviewed in Section \ref{sec:expbasics}. Section
\ref{sec:systematics} presents experimental results, with an emphasis on
describing the systematics of source dimensions and lifetimes as a function of
beam energy, system size, particle species and a particle's momentum. In
addition to source dimensions, results for phase space density and entropy are
presented. Comparisons of experimental results and transport models are
presented in Section \ref{sec:interpretations}, with an emphasis on explaining the
``HBT puzzle,'' i.e., the fact that dynamic descriptions that incorporate a
phase transition to a new state of matter with many degrees of freedom
significantly over-predict observed source sizes. Section \ref{sec:summary}
summarizes the current state of the field and lists new directions and
challenges for future theoretical and experimental analyses.

\section{Theory and Phenomenology Basics}
\label{sec:theoryBasics}

\subsection{Formalism}

Two-particle correlation functions are constructed as the ratio of the measured
two-particle inclusive and single-particle inclusive spectra,
\begin{eqnarray}
C^{ab}({\bf P},{\bf q}) &=&\frac{dN^{ab}/(d^3p_ad^3p_b)}
{(dN^a/d^3p_a)(dN^b/d^3p_b)}\, ,
\label{eq:FormalDefinition}
\\
\nonumber
P&\equiv& p_a+p_b,~~
q^\mu=\frac{(p_a-p_b)^\mu}{2}-\frac{(p_a-p_b)\cdot P}{2P^2}P^\mu\, .
\end{eqnarray}
The theoretical analysis of (\ref{eq:FormalDefinition}) aims at relating this
experimentally measured correlation to the space-time structure of the particle
emitting source
\cite{Heinz:1999rw,Bauer:1993wq,Wiedemann:1999qn,Tomasik:2002rx}. Two forms
are common for connecting the measured correlation function to the space-time
emission function $s(p,x)$ through a convolution with the wave function
$\phi$. In the first form \cite{Lednicky:1981su},
\begin{equation}
\label{eq:altmaster}
C^{ab}({\bf P},{\bf q})=\frac{
\int d^4x_a d^4x_b s_a(p_a,x_a) s_b(p_b,x_b) 
|\phi({\bf q}',{\bf r}')|^2}
{\int d^4x_a s_a(p_a,x_a) \int d^4x_b s_b(p_b,x_b)}\, .
\end{equation}
In calculations of two-particle correlation functions, the squared relative
two-particle wave function $|\phi|^2$ serves generally as a weight, and the
emission function $s(p,x)$ contains all space-time information about the source
because it describes the probability of emitting a particle with momentum $p$
from a space-time point $x$.  Here, and throughout this section, primes denote
quantities in the center-of-mass frame, i.e., the frame where ${\bf P}=0$.  The
source function $s_a$ is evaluated at the momentum $\bar{{\bf p}}_a=m_a{\bf
  P}/(m_a+m_b)$, $\bar{p}_a^0=E_a(\bar{\bf p}_a)$.  

The second form, which is equally justified as Equation~\ref{eq:altmaster} by the
approximations described further below, is,
\begin{eqnarray}
\label{eq:master}
C^{ab}({\bf P},{\bf q})-1&=&\int d^3 r' {\cal S}_{\bf P}({\bf r}') 
\left[|\phi({\bf q}',{\bf r}')|^2-1\right],\\
\nonumber
{\cal S}_{\bf P}({\bf r}')&\equiv& 
\frac{\int d^4x_ad^4x_b s_a(\bar{p}_a,x_a)
s_b(\bar{p}_b,x_b)
\delta\left({\bf r}'-{\bf x}_a^\prime+{\bf x}_b^\prime\right)}
{\int d^4x_ad^4x_b s_a(\bar{p}_a,x_a) s_b(\bar{p_b},x_b)}~.
\end{eqnarray}
This expression allows one to consider $|\phi|^2$ as a kernel with which one
can transform from the coordinate-space basis to the relative-momentum
basis. It also emphasizes the limitation of correlation functions, that they
can provide, at best, the function ${\mathcal S}_{\bf P}({\bf r}')$, the
distribution of relative positions of particles with identical velocities and
total momentum ${\bf P}$ as they move in their asymptotic state.  Thus,
correlations do not measure the size of the entire source. Instead, they
address the dimensions of the ``region of homogeneity,'' a term coined by
Sinyukov \cite{Akkelin:1995gh}, i.e., the size and shape of the phase space
cloud of outgoing particles whose velocities have a specific magnitude and
direction.  If the collective expansion of the produced matter is strong, as is
the case in central collisions, then the region of homogeneity is significantly
smaller than the entire source volume.  In the following, we discuss the
assumptions on which Equations~\ref{eq:altmaster} and~\ref{eq:master} are
based.

We start from explicit expressions for the one- and two-particle spectra in
terms of $T$-matrix elements. For one-particle emission,
\begin{eqnarray}
\label{eq:p1tmatrix}
E\frac{dN}{d^3p}&=&\int d^4x~s(p,x)
=\sum_{F'}|\int d^4x T_{F'}(x) e^{-ip\cdot x}|^2,\\
s(p,x)&=&\sum_{F'}\int d^4\delta x T_{F'}^*(x+\delta x/2)
T_{F'}(x-\delta x/2) e^{ip\cdot\delta x}\, .
\end{eqnarray}
Here, $F'$ refers to the state of all other particles in the system. All
interactions with the residual system are incorporated into the $T$ matrix.
However, there is a choice as to whether mean-field interactions are included
in the $T$ matrix or are instead incorporated into the evolution matrix
\cite{Barz:1998ce,Barz:1996gr,Cramer:2004ih,Chu:1994de,Kapusta:2005pt}. For
example, one can include the Coulomb interactions with the residual system by
replacing the phase factor $e^{ip\cdot x}$ in Equation~\ref{eq:p1tmatrix} with an
outgoing Coulomb wave function. This can be quantitatively important, in
particular for slow particles. It becomes difficult when the two particles
interact with one another through the strong or Coulomb force, as this
represents a quantum three-body problem.

{\it Assumption 1: Higher order (anti)symmetrization can be neglected.} Equation~\ref{eq:p1tmatrix} implies that all particles with
asymptotic momentum $p$ must have had their last interaction with the source at
some point $x$.  For distinguishable particles, this is indeed the case and
Equation~\ref{eq:p1tmatrix} does not represent an assumption. However, if there
are $N_a>1$ particles of the same type $a$, then one must consider
$T_a(x_1\cdots x_{N_a})$.  The evolution matrix is then no longer a simple
phase factor but includes $N_a!$ interference terms. The single-particle
probability is then obtained by integrating over the other $N_a-1$ momenta.
This can be accomplished explicitly for simple source
functions~\cite{Pratt:1993uy,Heinz:2000uf,Zhang:1998sz,Csorgo:1997us,Zimanyi:1997ur}.
The distortion to the single-particle spectra and to the two-particle
correlation function were found to be important when the phase space density
approached unity. Otherwise, Equation~\ref{eq:p1tmatrix} is well justified.

\begin{figure}[t!]
\centerline{\includegraphics[width=0.7\textwidth]{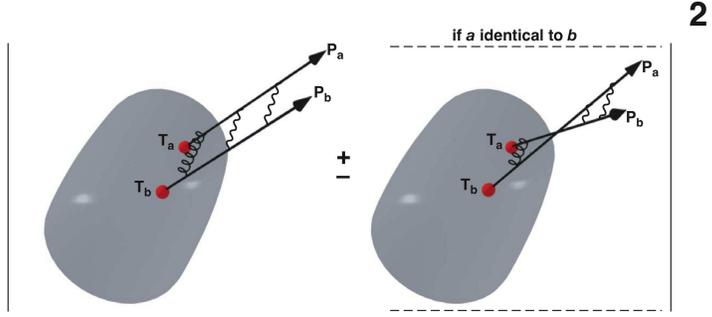}}
\caption{\label{fig:kopylov}Schematic representation of the squared emission
  amplitude for two particles emitted independently from the grey-shaded
  source region and interacting with each other in the final states. For
  identical bosons (+) and fermions (-), correlation also involves interference
  between the paths.}
\end{figure}

{\it Assumption 2: The emission process is initially uncorrelated.}  In writing Equation~\ref{eq:p1tmatrix}, one requires that
two-particle matrix elements factorize,
$T_{F''}(x_a,x_b)=T_{F'_a}(x_a)T_{F'_b}(x_b)$, i.e., that the emission is
independent. If multi-particle symmetrization can be neglected, the
two-particle evolution operator factorizes into a center-of-mass and a relative
operator. One has $U(x_a,x_b;p_a,p_b)=u_{q'}(x'_a-x'_b)
e^{iP\cdot(E'_ax_a/M_{\rm inv}+E'_bx_b/{\rm Minv})}$ for non-identical
particles, whereas for identical particles
$U=e^{iP\cdot(x_a+x_b)/2}(u_{q'}(x'_a-x'_b)
\pm(u_{q'}(x'_b-x'_a))/\sqrt{2}$. This is illustrated in
Figure~\ref{fig:kopylov}. Then, the two-particle probability can be expressed
in terms of one-particle source functions,
\begin{eqnarray}
&& E_aE_b\frac{dN_{ab}}{d^3p_ad^3p_b}
=\int d^4x_a d^4x_b d^4\tilde{q}
s_a((E'_a/M_{\rm inv})P+\tilde{q},x_a) s_b((E'_b/M_{inv})P-\tilde{q},x_b)
\nonumber \\
&& \qquad \qquad  
\times d^4\delta r' e^{i\tilde{q}\cdot\delta r'}
u_{q'}^*(x_a'-x_b'+\delta r'/2)u_{q'}(x_a'-x_b'-\delta r'/2).
\label{eq:intermediate}
\end{eqnarray}

{\it Assumption 3: Smoothness approximation \cite{Anchishkin:1997tb,Zhang:1997db,Pratt:1997pw}}. Equation~\ref{eq:intermediate} is difficult to evaluate as it requires knowledge of
the source function evaluated off-shell. For the special case where the
particles do not interact aside from identical particle interference,
$u_q(x_a-x_b)=[e^{iq\cdot(x_a-x_b)}\pm e^{iq\cdot(x_b-x_a)}]/\sqrt{2}$, the
integrals over $\tilde{q}$ and $\delta r'$ can be performed analytically,
\begin{eqnarray}
\label{eq:intermediate_hbt}
E_aE_b\frac{dN_{ab}}{d^3p_ad^3p_b}
&=&\int d^4x_a d^4x_b
\left\{s(p_a,x_a) s(p_b,x_b)\right.\\
\nonumber &&\left.\pm
s(P/2,x_a)s(P/2,x_b)\cos[(p_a-p_b)\cdot(x_a-x_b)]\right\}.
\end{eqnarray}
The source functions in the interference term are evaluated off-shell for
non-zero $q$, $P_0/2\ne E({\bf P}/2)$. The smoothness approximation replaces
$s(P/2,x_a)s(P/2,x_b)$ with either $s(E({\bf P}/2),{\bf P}/2,x_a) s(E({\bf
  P}/2),{\bf P}/2,x_b)$, which leads to Equation~\ref{eq:master}, or with
$s(p_a,x_a)s(p_b,x_b)$, which leads to Equation~\ref{eq:altmaster}. If the first
approximation is performed, one should also make the same approximation for the
denominator. The smoothness approximation has been checked for expanding
thermal sources, where it was found to be very reasonable for large (RHIC-like)
sources, but quite questionable for smaller sources such as those found in $pp$
or $e^+e^-$ collisions \cite{Pratt:1997pw}.

{\it Assumption 4: Equal time approximation.} For the general
case where the evolution operator incorporates Coulomb or strong interactions,
deriving Equations~\ref{eq:master} and~\ref{eq:altmaster} from
Equation~\ref{eq:intermediate} is more complicated. First, the smoothness
assumption amounts here to neglecting the $\tilde{q}$ dependence in the product
of the source functions in Equation~\ref{eq:intermediate}. This assumption is
somewhat more stringent in the presence of final state interactions because the
relevant range of $\tilde{q}$ extends beyond $q$. With this assumption, one
obtains a $\delta$-function constraint for $\delta r'$, and the integrand of
(\ref{eq:intermediate}) is proportional to the squared evolution matrix
$|u_q'(x_a'-x_b')|^2$. This evolution matrix has non-zero time components,
which must be neglected if one is to identify it with the relative wave
function. because the relative motion in the pair rest frame is small, one
expects this approximation to be reasonable, but it has not yet been tested in
model studies.

The above formalism is semi-classical in the sense that a quantum-mechanical
particle emission probability, defined by the $T$-matrix elements, is
usually approximated by classical source functions. As a consequence, quantum
uncertainty limits the applicability of Equations~\ref{eq:master} and~\ref{eq:altmaster}. To illustrate this limitation, source functions have
been evaluated by convoluting the emission function with wave
packets\cite{Aichelin:1996iu,Wiedemann:1997kq,Wiedemann:1998ng,Padula:1998ti}
of spatial width $\sigma$. This leads to a broadening of spatial distributions
by $\left(\Delta R\right)^2 \sim 0.5 - 1.0$ fm$^2$. because quantum
smearing may already be incorporated into some of the semi-classical treatments
through the choice of the initial density distribution, these calculations
should be regarded as indicative of the theoretical uncertainty. because this
error affects the size in quadrature, it is negligible for large sources, but
might be significant for sources near 1 fm in size with strong space-time
correlations~\cite{Zhang:1997db,Pratt:1997pw,Wiedemann:1997kq,Wiedemann:1998ng,Martin:1998ku}.          
In particular, analyses of $\pi\pi$ correlations
from $e^+e^-$ collisions, which result in source sizes of less than 1.0
fm\cite{Barate:1999gj,Abreu:1999xf}, are difficult to interpret in the above
formalism. 

\subsection{Identical-Particle Interference}
In the absence of strong and electromagnetic final state interactions, the wave
function of an identical particle pair in Equation~\ref{eq:master} becomes
\begin{equation}
|\phi({\bf q}',{\bf r}')|^2-1=\pm\cos(2{\bf q}'\cdot{\bf r}').
\end{equation}
The distribution of separations in coordinate space ${\cal S}_{\bf P}({\bf
  r}')$ can then be determined by performing a three-dimensional Fourier
transform of $C({\bf q}')-1$. It is instructive to consider the properties of
this inversion in more detail. The curvature of $C({\bf q})$ at ${\bf q} = 0$
can be related to the mean-square separation of the three-dimensional shape of
${\cal S}_{\bf P}({\bf r})$ (we neglect the ${\bf P}$ labels in $C$ and ${\cal  S}$),
\begin{equation}
 \left. \begin{array}{r}-\frac{d^2 C({\bf q})}{dq_i^\prime\, dq_j^\prime}\end{array} \right|_{q=0}
=\langle r_ir_j\rangle
=\int d^3r {\cal S}({\bf r}) r_ir_j.  
\label{curv}
\end{equation}
This relation has been useful to qualitatively illustrate the relation between
specific space-time information and specific features of the
correlator. However, applying the identity quantitatively requires careful
consideration of pions from longer-lived resonances which can dominate the
calculation of $\langle r^2\rangle$ if not accounted for.

\subsection{Correlations from Coulomb and Strong Interactions}
\label{subsec:coulstrong}

Compared to the case for non-interacting identical particles, where the
transformation between ${\cal S}_{\bf P}({\bf r})$ and $C({\bf P},{\bf q})$ is
a Fourier transform, analyzing the experimentally measured correlation function
with Equation~\ref{eq:master} to determine the source function is more
complicated.  Understanding the resolving power of the kernel $|\phi({\bf
  q},{\bf r}|^2$ requires a detailed understanding of the relative wave
function.  Once one averages over spins, the squared relative wave function is
a function of $q$, $r$ and $\cos\theta_{qr}$.  In relativistic collisions,
correlation analyses are usually confined to light, singly charged
hadrons. Coulomb-induced correlations are then weak and must be
analyzed at small $q$, where quantum effects become important ($qr/\hbar\sim
1$). The relative two-particle wave function in the presence of Coulomb
interactions can then be written as a function of $qr/\hbar$, $r/a_0$ and
$\cos\theta_{qr}$.
\begin{eqnarray}
\phi&=& \Gamma(1+i\eta) e^{-\pi\eta/2}e^{i{\bf q}\cdot{\bf r}}
\left\{1+\sum_{n=1}^\infty h_n \cdot(r/a_0)^n\right\},
\end{eqnarray}
where $a_0$ is the Bohr radius, $h_1 = 1$ and $h_n=\frac{n-1-i\eta}{-in\eta}\,
h_{n-1}$.  Here, $\eta\equiv\mu e^2/\hbar q$ is independent of $r$, and for
small $r/a_0$ the correlation function behaves as the Gamow factor,
$G(\eta)\equiv e^{-\pi\eta}|\Gamma(1+i\eta)|^2=2\pi\eta/(e^{2\pi\eta}-1)$.
Thus, the Coulomb kernels have little resolving power for $\pi\pi$ correlations
where $a_0=387$ fm, but have greater resolving power for $pK$
correlations where $a_0=84$ fm.

For $r<<a_0$,
the effects of Coulomb interactions can be removed easily from the correlation function
because $\eta$ is independent of $r$ \cite{Gyulassy:1979yi},
\begin{equation}
|\phi|^2\approx G(\eta)[1+\cos(2qr\cos\theta_{qr})]\, .
\end{equation}
For realistic source sizes, the order $r/a_0$ corrections are of the order of
10\% for $\pi\pi$ correlations and are larger for heavier pairs
\cite{Anchishkin:1997tb,Pratt:1986ev}. Significant effort has been invested in
``correcting'' experimental correlation functions to remove Coulomb effects to
all orders, but such corrections are model-dependent. The safest method for
determining the source function is either inverting the full kernel
\cite{Brown:2000aj,Brown:1999ka,Brown:1997ku,Chung:2002vk,Verde:2001md,
  Verde:2003cx} or fitting $C({\bf q})$ to some parameterized form for ${\cal
  S}({\bf r})$, which is convoluted with the full kernel. Neither of these
tactics are computationally prohibitive.

Strong interactions can also be exploited to provide size and shape
information. If the size of the source is much larger than the range of the
potential between the two particles, the kernel $(|\phi|^2-1)$ can be
determined entirely from knowledge of the phase shifts. Pairs that have a
resonant interaction are especially useful, because the resonance will lead to
a peak whose height is inversely proportional to the source volume, if
$qR>>\hbar$.  At small $q$ the kernel is determined by the scattering length,
$a$ \cite{Wang:1999bf}, and the height of the correlation at $q=0$ becomes
\begin{equation}
C({\bf q=0})-1=\left\langle -\frac{2a}{r} +\frac{a^2}{r^2} \right\rangle\, ,
\end{equation}
where the averaging is performed using ${\cal S}_{P}({\bf r})$ as a weight.  Of
course, the effects of strong interactions, Coulomb interference, and identical-particle
interference can all combine as is the case for $pp$ correlations. The $pp$
kernel has been analyzed in depth by Brown and Danielewicz, where the kernel was
inverted and applied to experimental $pp$ data. Evidence was seen for
significant non-Gaussian behavior in the resulting source functions
\cite{Verde:2001md,Verde:2003cx}. Strong interactions also provide angular
resolving power \cite{Pratt:2003ar} which can be understood from the
perspective of classical trajectories. Even $s$-wave scattering can be
exploited to discern information about shape.

Strong and Coulomb-induced correlations apply to both identical and
non-identical particles. For non-identical particles, the wave function is not
symmetrized and $|\phi({\bf q},{\bf r})|^2\ne |\phi({\bf q},-{\bf r})|^2$, which
results in odd components of the correlations function, $C({\bf q})\ne C(-{\bf
  q})$ if there are odd components of ${\cal S}({\bf r})$ as is the case for
non-identical particle pairs. This asymmetry can be experimentally exploited
to investigate the spatio-temporal differences between the emission functions
of different particle species
\cite{Lednicky:1995vk,Voloshin:1997jh}; this requires, however, sufficient
statistics to select on the angle between the total and relative momentum in
the pair center of mass $\angle({\bf q},{\bf P})$~\cite{Adams:2003qa}.

\subsection{Coordinate Systems}

Correlation functions depend on two three-dimensional momenta, ${\bf P}$ and
${\bf q}$. For high-energy collisions, analyses are usually performed in the
longitudinally comoving system (LCMS), a rest frame moving along the
longitudinal (beam) direction such that $P_z=0$. Axes are usually chosen
according to the out-side-long prescription. The longitudinal axis is
chosen parallel to the beam, while the outward axis points in the direction of
${\bf P}$, which is perpendicular to the beam axis. The sideward axis is chosen
perpendicular to the other two. If the system is boost-invariant, observables
expressed in the LCMS variables are independent of $P_z$ and the source has a
reflection symmetry about the $r_{\rm long}=0$ plane. If the collision is
central, there is also a reflection symmetry about the $r_{\rm side}=0$ plane.
Any four-vector $V$ can be expressed in this coordinate system using the
four-momentum $P$ to project out the
components\cite{Pratt:1986cc,Bertsch:1989vn,Csorgo:1989kq},
\begin{eqnarray}
V_{\rm long}&=&(P_0V_z-P_zV_0)/M_T,\\
\nonumber
V_{\rm out}&=&(P_xV_x+P_yV_y)/P_T,\\
\nonumber
V_{\rm side}&=&(P_xV_y-P_yV_x)/P_T,
\label{typicalvectors}
\end{eqnarray}
where $M_T^2=P_0^2-P_z^2$ and $P_T^2=P_x^2+P_y^2$. Dimensions of the source
function are typically quoted in this coordinate system. One could also perform
a second boost to the pair frame, in which the transverse components of the
total momentum are zero. Then,
\begin{equation}
V_{\rm out}'=\frac{M_{\rm inv}}{M_T}\frac{(P_xV_x+P_yV_y)}{P_T}
-\frac{P_T}{M_TM_{\rm inv}}P\cdot V,
\end{equation}
where $M_{\rm inv}^2=P^2$. Relative wave functions are more conveniently
expressed in the frame of the pair. For instance, a sharp resonant peak is no
longer sharp if the correlation is viewed away from the pair frame.  For pairs
where the correlation is influenced by Coulomb and strong interactions, most
analyses are conducted in the pair frame.

For non-zero impact parameters, azimuthal symmetry is lost and source functions
also depend on the azimuthal direction of the pair's total momentum
\cite{Wiedemann:1997cr,Lisa:2000ip,Heinz:2002au}. Also, if boost-invariance is
broken, the pair's rapidity needs to be specified. In this more general case,
reflection symmetries are broken and the choice of the coordinate axes becomes
somewhat arbitrary. One could orient the axes according the event's impact
parameter, or one could rotate the coordinate system so that in the new frame
cross terms such as $\langle xy\rangle$ vanish (illustrated in
Figure~\ref{fig:CrossTerm}). One would then specify the Euler angles as part of
the description of the source function.

\subsection{Gaussian Parameterizations}
\label{sec:GaussianParams}

To gain a physical understanding of the three-dimensional spatio-temporal
source distribution, it is useful to summarize its size and shape with a few
parameters. This motivates the study of Gaussian parameterizations for the
source ${\cal S}_{\bf P}({\bf r}')$ and the two-particle correlator.  Realistic
sources deviate from Gaussians, e.g. by exponential tails caused by resonance
decay contributions. The extracted Gaussian source parameters may thus depend
on details of the fitting procedures. These shortcomings can be overcome with
imaging methods, or more complicated forms for the fitting. A more general
three-dimensional analysis of correlations would involve decomposing both the
correlation and source functions in either spherical or Cartesian harmonics
\cite{Danielewicz:2005qh,ZibiMikeTom}. Although the detailed non-Gaussian aspects of the
correlation are important, the extra information can also cloud the main trends
in the data. In practice, Gaussian parameterizations provide the standard
minimal description of experimental data.

\subsubsection{The general case}

The most general form for a Gaussian source is
$\exp\{-A_{\alpha\beta}(x_\alpha-\bar{x}_\alpha)(x_\beta-\bar{x}_\beta)\}$,
where $x_\alpha$ refers to the four dimensions $x,y,z,t$, and $A$ is a $4\times
4$ real symmetric matrix. The most general form has 14 parameters, 10
parameters for $A$ and four more parameters for the offsets
$\bar{x}_\alpha$. Reflection symmetries can be used to eliminate certain cross
terms and some of the offsets \cite{Chapman:1995nz}. Furthermore, if the
particles are identical or have the same phase space distributions, all the
offsets can be set to zero. because the source function for the second species
$b$ might also have 14 parameters, there could be up to 28 Gaussian parameters
in describing both $s_a$ and $s_b$. However, because ${\cal S}({\bf r})$ depends
only on the distribution of relative spatial coordinates in the pair frame,
only nine Gaussian parameters are required to describe the most general ${\cal S}$ for a given ${\bf P}$. Three of these nine parameters can be
identified with Gaussian widths, three can be identified with offsets, and the
last three can either be identified with cross terms or with the three Euler
angles describing the orientation of the three-dimensional ellipse.

Using the reflection symmetries for mid-rapidity sources in a symmetric central
collision, a Gaussian parameterization of the emission function for particle
species $a$ in the out-side-long coordinate system reads
\begin{eqnarray}
s_a(p,x)&\sim&
\exp\left\{-\frac{(x_{\rm out}-\bar{x}_{a,{\rm out}}-V_{s,a}(t-\bar{t}_a))^2}
{2R_{a,{\rm out}}^2}\right. \\
\nonumber
&&\hspace*{48pt}\left.-\frac{(x_{\rm side})^2}{2R_{a,{\rm side}}^2}
-\frac{(x_{\rm long})^2}{2R_{a,{\rm long}}^2}
-\frac{(t-\bar{t}_a)^2}{2(\Delta\tau_a)^2}
\right\}.
\end{eqnarray}

The symmetries forbid any cross terms in the exponential involving $x_{\rm
  side}$ or $x_{\rm long}$, such as $x_{\rm side}x_{\rm out}$ or $x_{\rm
  long}t$, but do not forbid a cross term between outward position $x_{\rm
  out}$ and time $t$. Here, this cross term is taken into account by allowing
the source to move in the outward direction with a velocity $V_s$. The
symmetries also forbid offsets in the sideward or longitudinal direction. These
other offsets have also been addressed for non-central collisions
\cite{Retiere:2003kf,Miskowiec:1998ms}.

The correlation function is determined by the phase space density of the final
state, Equation~\ref{eq:bertschfbar}. The resulting phase space density is
\begin{eqnarray}
\label{eq:f_gauss}
\nonumber
f_a({\bf p},{\bf r},t)&\sim&\exp
\left\{
-\frac{\left[x_{\rm out}-\bar{X}_a(t)\right]^2}
{2\left[R_{a,{\rm out}}^2 +(V_{s,a}-V_\perp)^2(\Delta\tau_a)^2\right]}
-\frac{x_{\rm side}^2}{2R_{a,{\rm side}^2}}
-\frac{x_{\rm long}^2}{2R_{a,{\rm long}^2}}
\right\},\\
\bar{X}_a(t)&=&\bar{x}_{a,{\rm out}}+V_\perp(t-\bar{t}_a).
\end{eqnarray}
Here, $V_\perp$ is the velocity of the pair in the LCMS frame.

The correlation function is determined by the relative distance function ${\cal
  S}_{\bf P}({\bf r}')$, see Equation~\ref{eq:master},
\begin{eqnarray}
\label{eq:gausssummary}
{\cal S}_{\bf P}({\bf r}')&\sim& \exp\left\{
-\frac{\left[r'_{\rm out}-\bar{X}_{\rm out}\right]^2}
{4\gamma_\perp^2R_{\rm out}^2}
-\frac{r_{\rm side}^{\prime 2}}{4R_{\rm side}^2}
-\frac{r_{\rm long}^{\prime 2}}{4R_{\rm long}^2}
\right\}\\
\nonumber
R_{\rm out}^2&=&\frac{1}{2}\left[R_{a,{\rm out}}^2+R_{b,{\rm out}}^2
+(V_{s,a}-V_\perp)^2(\Delta\tau_a)^2+(V_{s,b}-V_\perp)^2(\Delta\tau_b)^2\right],\\
\nonumber
R_{\rm side}^2&=&\frac{1}{2}\left[R_{a,{\rm side}}^2
+R_{b,{\rm side}}^2\right],~~~
R_{\rm long}^2=\frac{1}{2}\left[R_{a,{\rm long}}^2
+R_{b,{\rm long}}^2\right],\\
\nonumber
\bar{X}_{\rm out}&=&\bar{x}'_{a,{\rm out}}-\bar{x}'_{b,{\rm out}},
\end{eqnarray}
where $\gamma_\perp\equiv(1-V_\perp)^{-1/2}$. Thus, there are four measurable
parameters, $R_{\rm out}$, $R_{\rm side}$, $R_{\rm long}$, and $\bar{X}_{\rm
  out}$. In the absence of any symmetry, there are five more terms: three cross
terms $(R^2_{\rm out,side}, R^2_{\rm out,long}$ and $R^2_{\rm side,long})$, and
two more offsets ($X_{\rm side}$ and $X_{\rm long}$). For identical particles,
all the offsets are zero.
\subsubsection{Sensitivity to Lifetime}
Given the symmetries used to derive Equation~\ref{eq:gausssummary}, the most
experiment can provide is the determination of the four parameters, $R_{\rm
  out}$, $R_{\rm side}$, $R_{\rm long}$, and $\bar{X}_{\rm out}$.  The only way
to independently determine the lifetime is to assume that the two transverse
spatial sizes are approximately equal \cite{Heinz:1996qu}. After assuming
$R_{a,{\rm out}}^2 + R_{b,{\rm out}}^2\approx R_{b,{\rm side}}^2+R_{b,{\rm
    side}}^2$, Equation~\ref{eq:gausssummary} yields for identical particles
($V_{a,s}=V_{b,s}=V_s,\Delta\tau_a=\Delta\tau_b=\Delta\tau$),
\begin{equation}
(V_\perp-V_s)^2(\Delta\tau)^2\approx R_{\rm out}^2-R_{\rm side}^2\ .
\label{eq:emissionduration}
\end{equation}
In general, however, the source velocity is not precisely known and outward and
sideward spatial dimensions are not exactly equal; these factors result in a significant
systematic error when applying Equation~\ref{eq:emissionduration}, especially
because temporal effects enter in quadrature. The $R_{\rm out}/R_{\rm side}$
ratio only provides a reliable estimate of the lifetime when
$(V_\perp-V_s)\Delta\tau$ is much larger than the transverse size.

\subsubsection{Gaussian Cross Terms}
\label{sec:CrossTerms}
\begin{figure}[t!]
\centerline{\includegraphics[width=0.7\textwidth]{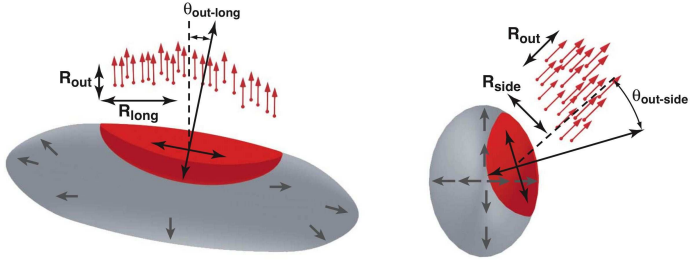}}
\caption{\label{fig:CrossTerm} For non-central collisions the principal axes
  describing the orientation of the region of homogeneity can differ from the
  out-side-long axes. By viewing the source distribution from the perspective
  where the beam axis is oriented horizontally ({\it left panel}) and from peering
  down the beam pipe ({\it right panel}), the orientations leading to out-long and
  out-side cross terms are illustrated.}
\end{figure}

In addition to the three Gaussian parameters, $R_{\rm out}$, $R_{\rm side}$ and
$R_{\rm long}$, that describe the dimensions of a Gaussian source, one needs, in
the general non-symmetric case, three more parameters to describe the angular
orientation of the principal axes. Figure~\ref{fig:CrossTerm} displays how the
principal axes might differ from the out-side-long axes once the collisions are
off center. These three Euler angles combined with three sizes can also be
related to the parameters $A_{ij}$ in the general form for a three-dimensional
Gaussian, $\exp (-A_{ij}r_ir_j)$, where $A$ is a symmetric matrix with six
independent components
\cite{Voloshin:1995mc,Chapman:1995nz,Wiedemann:1997cr,Heinz:2002au}.

For pairs of identical particles, the correlation function for
Gaussian sources is also Gaussian,
\begin{equation}
C({\bf q})=1+\exp(-({\cal R}^2)_{ij}Q_iQ_j)\, ,
\label{eq:C2Gauss}
\end{equation}
where $Q=2q$. The six experimentally determined parameters, $({\cal
  R})^2_{ij}$, can be related to the moments of ${\cal S}({\bf r}')$
\cite{Wiedemann:1996ej},
\begin{equation}
\langle r_ir_j\rangle =({\cal R}^2)_{ij}.
\end{equation}

For central collisions, the source sizes $(R^2)_{ij}$ depend only on the
longitudinal pair momentum $P_L$ and on the modulus of the transverse 
pair momentum $|\vec{P}_T|$. This is different for non-central collisions for which
the azimuthal direction $\phi_{\rm pair} = \angle(\vec{P}_T,\hat{b})$ of the 
transverse pair momentum with respect to the impact parameter direction $\hat{b}$ matters. The $\phi_{\rm pair}$-dependences of
$(R^2)_{\rm out}$,  $(R^2)_{\rm side}$, and $(R^2)_{\rm out-side}$
then characterize the degree to which the initially out-of-plane-extended source
geometry has expanded to the point where it becomes in-plane-extended
\cite{Wiedemann:1997cr,Lisa:2000ip}. The out-longitudinal and side-longitudinal cross terms
contain information about the extent to which the main axis of the emission
ellipsoid is tilted with respect to the beam axis\cite{Lisa:2000ip}.
We return to this topic in Section~\ref{sec:systematics-phi}.

The Yano-Koonin
parameterization~\cite{Yano:1978gk,Wu:1996wk,Chapman:1995nz,Chapman:1994yv}
provides an alternative basis for describing the out-long cross term.  The
Yano-Koonin form is based on the assumption that one can boost along the beam
axis to a source frame where the correlation function has a simple form.
\begin{equation}
\label{eq:yk}
C({\bf P},{\bf Q})=1+\exp\left\{
-\tilde{Q}_\perp^2R_\perp^2-\tilde{Q}_{||}^2R_{||}^2-\tilde{Q}_0^2R_0^2
\right\}\ ,
\end{equation}
where ${\tilde Q}$ is the momentum difference defined in the source frame which
has rapidity $y_{\rm YK}$. In that frame $R_0$ is the Gaussian lifetime and
$R_\perp$ and $R_{||}$ are the dimensions of the source. This can be transposed
to the out-long-side frame by boosting $\tilde{Q}$ along the beam axis to the
frame where $P_z=0$, i.e. the LCMS frame, then using the fact that $Q_0=Q_{\rm
  out}P_\perp/P_0\equiv Q_{\rm out}V_\perp$ in the new frame. This yields
$\tilde{Q}_0=\cosh(y_{\pi\pi}-y_{YK}) Q_{\rm out} V_\perp
-\sinh(y_{\pi\pi}-y_{\rm YK})Q_{\rm long}$, and
$\tilde{Q}_{||}=\cosh(y_{\pi\pi}-y_{YK}) Q_{\rm long} -\sinh(y_{\pi\pi}-y_{\rm
  YK})Q_{\rm out} V_\perp$, where $y_{\pi\pi}$ is the rapidity of the LCMS
frame. Substituting these expressions into Equation~\ref{eq:yk} yields a cross
term in the exponential equal to $Q_{\rm out}Q_{\rm long}(R_0^2+R_{||}^2)\sinh
2(y_{\pi\pi}-y_{\rm YK})$, which disappears when $y_{YK}=y_{\pi\pi}$. By
fitting $y_{YK}$ as a function of the pair rapidity, aspects of boost
invariance can be tested. Given that the distribution of source rapidities
should fall off for large rapidities, one expects $y_{YK}$ to lag $y_{\pi\pi}$,
because pions of a given rapidity would more likely have been emitted from
sources with smaller rapidities \cite{Wu:1996wk}.

\subsection{The $\lambda$ factor}
\label{sec:lambda}

Many pions measured in experiment come from long-lived decays. Pions from weak
decays may or may not, depending on the experiment, be identified and removed
from the analysis, because their decay vertices are typically a few centimeters
from the reaction center. Decays from $\eta$ or $\eta'$ mesons occur a few
thousand fm away from the center of the collision. At these distances, they are
effectively uncorrelated with other particles but cannot be identified with
experiment. If a fraction $\lambda$ of the pairs originate from the
spatio-temporal region relevant for correlations, the correlation is muted by
the factor $\lambda$ \cite{Gyulassy:1979yi}. If the source function is divided
into two contributions, ${\cal S}({\bf r})=\lambda{\cal S}_{\rm local}+
(1-\lambda){\cal S}_\infty$, where both ${\cal S}_{\rm local}$ and ${\cal
  S}_\infty$ integrate to unity, the resulting correlation is
\begin{equation}
C({\bf q})=(1-\lambda)+\lambda\int d^3 r' {\cal S}_{\rm local}({\bf r}') 
\left[|\phi({\bf q}',{\bf r}')|^2-1\right].\\
\end{equation}
If the experimental sample includes a contamination from weak decays, $\eta$,
or mis-identified particles of a fraction $f$, the lambda factor becomes
$(1-f)^2$. It is not uncommon for this contamination factor to be near 30\%,
which results in $\lambda$ near one half.

Certainly, this division is somewhat artificial, as there are non-Gaussian
tails, or halos \cite{Csorgo:1994in}, to ${\cal S}_{\rm local}$ due to such
causes as the exponential fall-off of the source function in the longitudinal
direction or semi-long-lived resonances such as the $\phi$, whose lifetime is 40
fm/c.  Non-Gaussian behavior is a subject for either imaging
\cite{Brown:2000aj,Brown:1999ka,Brown:1997ku,Chung:2002vk,Verde:2001md,
  Verde:2003cx} or for more complicated parameterizations.

The $\lambda$ factor is often referred to as an incoherence factor, the name
being motivated by the properties of a coherent state, $\exp\{i\int d{\bf p}
\eta({\bf p})[a({\bf p})+a^\dagger ({\bf p})]\} |0\rangle$, which for identical
particles leads to no correlation. Coherent states represent highly correlated
emissions caused by phase coherence and thus violate the assumption of incoherence
or uncorrelated emission implied by Assumption 2 as described early in this section. The question of whether an observation of $\lambda\ne 1$ is due to coherence or due to contamination
from particles from far outside the source volume can be tested by analyzing
three-particle correlations \cite{Heinz:1997mr,Heinz:2004pv}. Such analyses
of data at both SPS and RHIC have been consistent with the incoherence
conjecture \cite{Boggild:1999tu,Bearden:2001ea,Adams:2003vd}.
Microscopic model calculations at the AGS reproduce the excitation function of
$\lambda$ when resonances contributions are included~\cite{Lisa:2000no}.

\subsection{Collective Flow and Blast-Wave Models}

Both longitudinal and radial collective expansion reduce the size of the
region of homogeneity, i.e., the relevant volume for particles of a given
velocity. For an infinite volume, the size of this region is set by the length
one must move before collective velocity overcomes the thermal velocity, $R\sim
V_{\rm therm}/(dv/dz)$.

\subsubsection{Longitudinal Flow}
\label{sec:TheoryLongitudinalFlow}

The Gaussian lifetime $\Delta\tau$ described in Equation~\ref{eq:emissionduration}
represents the spread of the emission times. A small value of $\Delta\tau$ does
not imply that particles were emitted early, but that they were emitted
suddenly. Inferring the mean time at which particles are emitted requires a
different assumption.  For instance, at RHIC, the initial nuclei are Lorentz
contracted by a factor of 100, and if there were no subsequent expansion,
$R_{\rm long}$ would be less than a Fermi. If one assumes that the system
expands along the beam axis with no longitudinal acceleration, the collective
velocity becomes
\begin{equation}
\label{eq:vcoll}
V_{{\rm coll},z}=z/t.
\end{equation}
If emission then comes from sources moving over a large range of rapidities (a
boost-invariant expansion), the dimension along the beam axis for the source
emitting zero-rapidity particles is determined by the distance one can move
before the collective velocity overwhelms the thermal velocity to force the
emission function back to zero. The size can then be expressed as:
\begin{equation}
\label{eq:meantauestimate}
R_{\rm long} \approx \frac{V_{\rm therm}}{dv/dz}
=V_{\rm therm} \langle t\rangle.
\end{equation}
Whereas $R_{\rm out}/R_{\rm side}$ gives information about the suddenness of
emission, $R_{\rm long}$ provides insight into the mean time at which emission
occurs given an estimate of the thermal velocity.

\begin{figure}[t!]
\centerline{\includegraphics[width=0.7\textwidth]{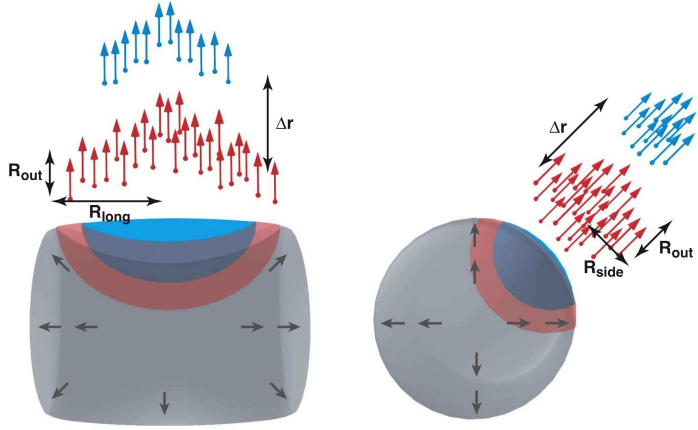}}
\caption{\label{fig:RvsMt} because particles with heavier masses have smaller
  thermal velocities, their source volumes are more strongly confined by
  collective flow. For longitudinal flow ({\it left panel}) this results in smaller
  values of $R_{\rm long}$ for particles with higher
  $m_T=\sqrt{m^2+p_T^2}$. For radial flow ({\it right panel}) this confines heavier
  particles toward the surface, which results in both a reduced volume and an
  offset $\Delta r$ in the outward direction.}
\end{figure}

For a thermal source with relativistic motion, the thermal velocity along the
beam axis is determined by the temperature and the transverse mass,
$m_T=\sqrt{m^2+p_T^2}$ \cite{Pratt:1986cc}. For large $m_T$ the thermal
velocity in the longitudinal direction becomes non-relativistic, $V_{\rm
  therm}=\sqrt{T/m_T}$, and the source size falls as $1/\sqrt{m_T}$ which is
referred to as $m_T$ scaling \cite{Makhlin:1987gm}. This is illustrated in
Figure~\ref{fig:RvsMt}. However, this assumes all particles are emitted with the
same Bjorken time $\tau_B$ and temperature, independent of the transverse
mass. because particles with high $m_T$ are probably emitted at lower $\tau_B$,
and because the temperature roughly behaves at $\tau_B^{-4/3}$, the longitudinal
size could fall even more quickly than $m_T^{-1/2}$.

In a boost invariant expansion, emission is a function of the Bjorken time
$\tau_B=\sqrt{t^2-z^2}$, not the time $t$, and because $t=\sqrt{\tau_B^2+z^2}$,
those particles emitted with small $z$ have a head start. This is sometimes
referred to as an inside-outside cascade. The transverse shape of ${\cal
  S}({\bf r})$ is then affected non-trivially by the expansion along the
beam. The resulting correlation function can be calculated analytically in the
case of pure identical-particle correlations
\cite{Kolehmainen:1986fe,Padula:1989ie}.

Boost invariance is incorporated
into blast-wave models with transverse expansion and assumed for many hydrodynamic models. The finite size of the system would alter
the results for two reasons. First, if the distribution of sources covers only
a finite range in $\eta$, the tails of the distribution ${\cal S}({\bf r})$ are
chopped off. Assuming the distribution in $\eta$ is Gaussian rather than
uniform,
\begin{equation}
\frac{1}{R_{\rm long}^2} \sim \frac{1}{V_{\rm therm}^2\tau_B^2}
+\frac{1}{\eta_G^2\tau_B^2},
\end{equation}
where $\eta_G$ is the range of rapidities over which the sources are
distributed. If $\eta_G$ were 1.5 units of rapidity, the extracted values of
$\tau_B$ from boost-invariant pictures would be underestimating $\tau_B$ by
$\sim 10$\%.

A second shortcoming of boost-invariant models is that they ignore acceleration
in the longitudinal direction. Accounting for this acceleration would alter the
relation between the time and the velocity gradient. Neglecting this
acceleration could also lead to a modest underestimate of $\tau_B$~\cite{Renk:2003gn}. 


\subsubsection{Transverse Collective Flow}
\label{sec:transverseFlow}

because transverse collective flow is intimately related to the pressure and
viscosity, it is of central importance. Blast-wave models are based on pictures
of thermal sources superimposed onto the transverse and longitudinal
collective velocity profiles. Simple forms are then chosen for the
profiles. Only two parameters are important for analyzing spectra, the
temperature and the transverse velocity. because heavier particles are more
sensitive to flow than are light particles, the two parameters can be adjusted
to fit the spectra of several species.

In addition to the temperature and transverse velocity, correlation
measurements are also sensitive to the space-time parameters of the blast
wave. In a minimal parameterization this would include the lifetime $\tau_B$
and the transverse size $R$. More sophisticated models would also include a
spread in lifetime $\Delta \tau$ and a surface diffuseness $\Delta
R$. Additional parameters ensue when one considers sensitivity to the reaction
plane. Then, two parameters are needed to describe the transverse size, and two
parameters are required to describe the transverse collective velocity. These
parameters can then depend on the azimuthal direction of ${\bf P}$
\cite{Heinz:2002sq}.

Choosing a blast-wave parameterization involves a number of choices about the form of the parameterization \cite{Tomasik:2001uz}. Chemical
potentials and temperatures might be chosen to vary with the transverse
position $r$ \cite{Chojnacki:2004ec,Csorgo:2001xm} or might be chosen to be
uniform. A wide variety of parameterizations have been employed for the
transverse velocity profile, which might choose linear profiles for either $v$,
$\gamma v$ or the transverse rapidity $asinh(\gamma v)$. In some
parameterizations, the velocity profile has been chosen to rise quadratically
with $r$ \cite{Lee:1990sk,Schnedermann:1993ws}. Although hydrodynamics has been
invoked as justification for different parameterizations, profiles from
hydrodynamics vary according to the equation of state.

For particles moving much faster than the surface velocity, transverse flow
manifests itself by constraining particles to an increasingly small fraction of
the blast-wave volume for the same reason that $R_{\rm long}$ falls with $m_T$
owing to longitudinal expansion \cite{Pratt:1984su}. For large $m_T$, this leads
to both $R_{\rm side}$ and $R_{\rm out}$ falling as $1/\sqrt{m_T}$
\cite{Chapman:1995nz,Chojnacki:2004ec,Csorgo:2001xm,Sinyukov:1998ex}. The fact that transverse
dimensions fall with $m_T$ might also result from the dynamics of cooling,
superimposed with a growing fireball. This correlates high-energy particles
with earlier times when the fireball was both smaller and at a higher
temperature.

Non-identical particles are of special interest in a blast-wave. For particles
moving faster than the surface of the blast wave, there is a stronger tendency
for heavier particles to be more confined to the region of the surface owing to
their slower thermal velocities \cite{Retiere:2003kf}. This results in heavy
particles being ahead of lighter particles of the same asymptotic velocity, and
leads to a non-zero $\Delta{\bf r}$ illustrated in Figure~\ref{fig:RvsMt}. As
discussed in Section \ref{subsec:coulstrong}, these displacements are accessible
through measuring odd components of the correlation function
\cite{Lednicky:1995vk,Voloshin:1997jh,Adams:2003qa}.

\subsection{Generating Correlations Functions from Hydrodynamics and 
from Microscopic Simulations}

Any model that predicts final-state space-time and momentum information of
emitted particles can be used to predict correlation functions. This
information may be extracted from both microscopic simulations or from
hydrodynamic calculations.

Microscopic simulations model the collision by evolving particles along
straight-line classical trajectories which are punctuated by collisions that
are programmed to be consistent with free-space cross-sections. When the
modeling is done on a one-to-one basis, the simulations are referred to as
cascades. Boltzmann simulations are similar but employ an oversampling by a
factor $N_s$ accompanied by a scaling down of the cross-sections by the same
factor. These are then consistent with the Boltzmann equation and become local
and relativistically covariant in the large $N_s$ limit
\cite{Molnar:2000jh,Cheng:2001dz}. To generate correlation functions from
either class of simulation, there are essentially two methods which are equally
justified within the smoothness approximation. Method I is motivated by
Equation~\ref{eq:master}. This involves first creating two lists, one for each
species, of the space time coordinates $x_a$ and $x_b$ of all those particles
that were emitted with momenta $m_a{\bf P}/(m_a+m_b)$ and $m_b{\bf
  P}/(m_a+m_b)$.  From these lists, one generates ${\cal S}_{\bf P}({\bf r})$
by sampling the distributions of $x_a-x_b$. This list is then convoluted with
$|\phi({\bf q},{\bf r})|^2$ to generate $C({\bf P},{\bf q})$ for all ${\bf
  q}$. In Method II, one samples pairs randomly without regard to their
momenta.  The numerator of the correlation function is then calculated by
generating pairs with the same weight as one expects to observe experimentally
and applying a weight given by the square of the relative wave function. The
denominator would be calculated in a similar manner, but without the weight
from the wave function. This method reflects the description of the correlation
function in Equation~\ref{eq:altmaster}. Acceptance effects or kinematic cuts can
then be performed exactly as they would be performed for real particles. Method
II has an advantage in that it is easier to accurately incorporate acceptance
effects or tight kinematic cuts. Method I makes for a much quicker calculation
because the procedure does not require sampling particles for irrelevant momenta
\cite{Zhang:1997db}.

Given the equation of state and the initial energy density, hydrodynamics
provides the means for solving for the space-time development of the
stress-energy tensor which can be used to make predictions for correlations
\cite{Kolb:2003dz,Huovinen:2001cy,Hirano:2004ta}. Viscous effects can also be
incorporated and are non-negligible
\cite{Heinz:2001xi,Teaney:2003pb}. Generating source functions from the output
of hydrodynamic calculations is not as straightforward as it might seem. The
Cooper-Frye prescription \cite{Cooper:1974qi} conserves energy and momentum if
the equation of state is one of free particles, but it suffers from the fact that
the particles that cross backwards across the surface into the hydrodynamic
volume enter the source function as a negative emission probability.  If the
relative velocity of the surface, as measured by an observer in the matter's
rest frame, is not much faster than the thermal velocity, a different
prescription is required. Numerous prescriptions have been proposed to address
these issues \cite{Tomasik:2002qt,Csernai:2004pr,Sinyukov:2002if}.

Hydrodynamic models, even those that incorporate viscosity, cannot be
justified once the system expands to the point that the mean free path is
similar to the characteristic size of the system. However, Boltzmann
descriptions or cascades are well justified at lower densities. Several efforts
have thus focused on coupling the two approaches
\cite{Soff:2000eh,Bass:2000ib,Teaney:2001gc,Teaney:2001av}. because the
final-state trajectories are established in the Boltzmann part of the
prescription, one can apply either of the methods mentioned above.

\subsection{Phase Space Density, Entropy and Coalescence}
\label{subsec:psdentropy}

because phase space density depends on both the momentum ${\bf p}$ and the
position ${\bf r}$, a measurement of the phase space density must specify the
spatial region over which it is determined. In practice, spatial information
from two-particle correlation functions is instrumental to this end. For
identical particle pairs, the ``area'' under the correlation function
determines the average phase space density \cite{Bertsch:1994qc}. Substituting
the final phase space density for the time-integrated source function,
\begin{equation}
\int^{t_f} \frac{dx_0}{E_p(2\pi)^3} s(p,x)|_{p_0=E_p}
=f({\bf p},{\bf x},t_f)\, ,
\end{equation}
and inserting into Equation~\ref{eq:master} with $Q=2q$ leads to
\begin{eqnarray}
\label{eq:bertschfbar}
\int d^3Q \left[C({\bf Q})-1\right] &=& (2\pi)^3
\frac{\int d^3r |f({\bf P}/2,{\bf r},t_f)|^2}
{\left[\int d^3r f({\bf P}/2,{\bf r},t_f)\right]^2}
= \frac{\bar{f}({\bf P}/2)}{dN/d^3p}\,  ,\\
\nonumber
\label{eq:fbardef}
\bar{f}({\bf p}) &=& \frac{(2\pi)^3}{(2S+1)}
\frac{E_p}{m}\frac{dN}{d^3p}{\cal S}_{{\bf P}=2{\bf p}}({\bf r}'=0).
\end{eqnarray}
Equation~\ref{eq:fbardef} applies also for the case of non-identical
particles of the same phase space density. We note that $\bar{f}({\bf p})$ is
the phase space density averaged over coordinate space for a specific momentum
using the phase space density itself as the weight. Unless $f({\bf p},{\bf
  r},t)$ is a constant within a fixed volume, the average phase space density
will fall below the maximum phase space density \cite{Tomasik:2002qt}. For
instance, if $f({\bf p},{\bf r},t)$ has a Gaussian profile in coordinate space,
the average phase space density will be $2^{-3/2}$ of the maximum phase space
density for that momentum. For a Gaussian source,
\begin{equation}
\label{eq:fbargauss}
\bar{f}({\bf p})=\frac{\pi^{3/2}}{(2S+1)mR_{\rm inv}^3}E\frac{dN}{d^3p},
\end{equation}
where $R_{\rm inv}^3=(E/m)R_{\rm out}R_{\rm side}R_{\rm long}$ is the product
of the three radii as measured in the frame of the pair.  The phase space
density is determined by combining a source size measurement with the spectra.
Entropy can be related to the phase space density in the standard way
\cite{Pal:2003rz}
\begin{eqnarray}
S&=&(2S+1)\int \frac{d^3rd^3p}{(2\pi)^3}
\left[-f\ln f\pm(1\pm f)\ln(1\pm f)\right],\\
\label{eq:dsdy}
dS/dy&\approx&\int d^2p_T ~E\frac{dN}{d^3p}\left[
\frac{5}{2}-\ln(2^{3/2}\bar{f}({\bf p})) \pm\bar{f}({\bf p})/2\right]\, .
\end{eqnarray}
Here, Equation~\ref{eq:dsdy} ignores higher powers of $\bar{f}$. 

The average phase space density is also straightforward to determine by
constructing ratios of spectra with species that can either bind or form a
resonance. If species $a$ and $b$ can bind to form species $c$, thermal
arguments would state
\begin{equation}
\label{eq:fprodthermal}
f_c({\bf r},t)=f_a({\bf P}m_a/m_c,{\bf r},t)
f_b({\bf P}m_b/m_c,{\bf r},t)e^{B/T},
\end{equation}
where $B$ is the binding energy, or the excitation energy if the resonance is
unstable. Coalescence arguments, which give the same expression but without the
binding energy \cite{Danielewicz:1991dh,LlopeCoalescence}, are identical if the
binding energy is small compared to the temperature, as is the case for nucleon
coalescence.  The average phase space density for $a$ or $b$ can be determined
by inserting Equation~\ref{eq:fprodthermal} into Equation~\ref{eq:fbardef},
\begin{equation}
\label{eq:fbarcoal}
\bar{f}_{a,b}({\bf p})=e^{-B/T}\frac{m_{b,a}(2S_{b,a}+1)}{m_c(2S_c+1)}
\frac{E_cdN_c/d^3P}{E_{b,a}dN_{b,a}/d^3p}.
\end{equation}
Here, the binding energy needs to be expressed in the frame of the thermal
bath. For the case where $B$ is small, the assumption of a thermal bath can be
neglected. Two examples where particles with similar phase-space densities form
low-energy resonances or bound states are $pn\rightarrow d$ and
$\phi\rightarrow K^+K^-$.

By comparing the expression for $\bar{f}$ from the ratio of spectra in
Equation~\ref{eq:fbarcoal}, with Equations~\ref{eq:fbardef} or~\ref{eq:fbargauss},
one can determine either ${\cal S}_{\bf P}({\bf r}'=0)$ or the Gaussian
parameters from ratios of spectra.
\begin{eqnarray}
{\cal S}_{\bf P}({\bf r}'=0)&=&
e^{-B/T}\frac{m_am_b(2S_a+1)(2S_b+1)}{(2\pi)^3m_c(2S_c+1)}
\frac{E_cdN_c/d^3P}{E_adN_a/d^3p_a\cdot dN/d^3p_b},\\
R_{\rm inv}^3({\bf P})&=&\frac{1}{(4\pi)^{3/2}{\cal S}_{\bf P}({\bf r}'=0)}.
\end{eqnarray}
Coalescence analyses can provide powerful measurements of volumes, but they
only provide a single number, ${\cal S}({\bf r}'=0)$, and they cannot provide any
insight into either the shape or the $r$ dependence.

\section{Femtoscopic measurements}
\label{sec:expbasics}

Experimental techniques have developed considerably in response to significant
improvements in both the theory and the quantity and quality of experimental
data.  In this section we discuss the general experimental approach for
defining and analyzing femtoscopic correlations and their systematic dependence
on global and kinematic quantities.

\subsection{Correlation Function Definition}

In practice, the formal definition of the correlation function in
Equation~\ref{eq:FormalDefinition} is seldom used in heavy ion physics.  Instead the
correlation of two particles, $a$ and $b$ for a given pair momentum ${\bf P}$
and relative momentum ${\bf q}$, is nominally given by
\begin{equation}
\label{eq:ExperimentalDefinition}
C^{ab}_{\bf P}({\bf q}) = \frac{A^{ab}_{\bf P}({\bf q})}{B^{ab}_{\bf P}({\bf q})}\cdot \xi_{\bf P}({\bf q}),
\end{equation}
where ${A^{ab}_{\bf P}({\bf q})}$ is the signal distribution, ${B^{ab}_{\bf
    P}({\bf q})}$ is the reference or background distribution which is ideally
similar to $A$ in all respects except for the presence of femtoscopic
correlations, and $\xi_{\bf P}({\bf q})$ is a correction factor introduced to
compensate for non-femtoscopic correlations present in the signal that are not
fully accounted for in the background as well as artifacts resulting, e.g., from
finite resolution and contamination.

\subsection{Signal Construction}

The signal ${A^{ab}_{\bf P}({\bf q})}$ refers to the relative momentum
distribution of particles $a$ and $b$ for a given range of pair momenta, ${\bf
  P}$, and a given set of event characterizations.  Although not all analyses
proceed in exactly the following fashion, the mechanics of constructing the
signal and background are most easily understood if one considers as separate
steps:
\begin{enumerate}
\item Event quality cuts and event-class binning;
\item Single-track (including particle identification) cuts and single-particle binning; and
\item two-particle pairing, two-track cuts, and pair momenta binning.
\end{enumerate}
Here the term event class refers to both physics observables, such as collision
centrality and reaction plane orientation, and detector considerations, such as
event vertex position and the condition of the detector when the event was
recorded, usually keyed by run number.  The latter considerations are relevant
to the proper construction of the background.  For event-class, particle, and
pair bin, the final signal is usually stored as a set of 3D-histograms in the
canonical relative momentum variables.

Single-particle acceptances divide out with a properly constructed background,
but 2-track acceptances can have a large effect on the correlation function.
For this reason the analysis of 2-track cuts is the dominant consideration in
the signal construction for most analyses.  The cuts and terminology are
different for Time Projection Chamber (TPC) experiments with near continuous hit distributions and Drift
Chamber experiments with projective geometry, but the goals are the same.
Split-tracks~\cite{Adams:2004yc,Lisa:2005vw} and ghost-tracks both refer to
single tracks which are incorrectly reconstructed as a pair of tracks with very
low relative momenta.  Even after the event-reconstruction algorithms (which
generate a list of individual tracks) have been optimally tuned, small traces
of these false pairs remain and must be removed from the analysis with
identical pairwise cuts.  Usually only a tiny fraction of tracks are
split, and this effect may be ignored in essentially all experimental analyses except
femtoscopic ones.  Various methods have been developed for identifying likely
split tracks, usually based on the number~\cite{Lisa:2005vw} or
topology~\cite{Adams:2004yc} of space-points associated with the track.

Pairwise effects usually also result in the loss of pairs at low relative
momentum, because two tracks with very similar trajectories tend to be
reconstructed as a single track.  (Note that in tracking detectors, this is not
a problem if one or both of the particles is a topologically identified neutral
particle.  In that case, the decay daughters may be well-separated even if
their parents have identical momentum.)  Such merging issues are usually
resolved by pairwise cuts that remove merged pairs.  Developing efficient and
appropriate cuts can be a subtle exercise, and it requires good knowledge of the
detector and event reconstruction software.  In most cases these cuts are
supported by simulations, but final determinations are nearly always based upon
data.

What does it mean to cut out merged pairs?  After all, if
the tracks have merged, then the pair is lost anyway.  The point is twofold.
First, the pair efficiency usually does not drop from 100\% to 0\% sharply as
a function of any variable.  Thus, the cuts are usually tuned to exclude
all~\cite{Lisa:2000no,Adams:2004yc} or
most~\cite{Boggild:1994vk,Ahle:2002mi,Adler:2004rq} of the inefficient region.
If regions with less than perfect efficiency remain in the analysis, a
2-track efficiency correction based on Monte Carlo simulations must be
applied, typically leading to systematic uncertainties of a few percent.  The
second reason for the cut is that it is applied equally to the signal and to
the background distribution $B^{ab}_{\bf P}({\bf q})$~\cite{Boggild:1993zj}.
Thus, if some fraction $f$ of pairs is lost at some relative momentum ${\bf q}$
in $A^{ab}_{\bf P}({\bf q})$, the same fraction is lost in $B^{ab}_{\bf P}({\bf
  q})$, and the ratio in Equation~\ref{eq:ExperimentalDefinition} is robust
against the effect.

\subsection{Background Construction}

For reasons described above, all cut-imposed effects on the signal pair distribution
$A^{ab}$ must be applied to $B^{ab}$.  This often means identifying which pairs would have been removed by merging, splitting, or other cuts, had the particles come from in the same event.

The ideal background should be identical to the signal in all respects except
for the presence of femtoscopic correlations.  Therefore, the global event
characteristics, single particle distributions, and acceptances should match
those of the signal.  A simple and straightforward way to construct such a
background is to form pairs from different events within a single event class.
This event-mixing technique~\cite{Kopylov:1974th} has gained wide acceptance in
relativistic heavy-ion collisions where violations to energy-momentum
conservation are negligible in the high multiplicity environment.  This
technique will be described in detail in what follows.  However, other methods
have also been used, especially if one considers femtoscopy in other systems.

For elementary-particle collisions or in low-multiplicity events, 
event mixing can violate total energy-momentum
conservation, especially when exclusive final states or jet-axes must be preserved;
thus, the correlation function would reflect non-femtoscopic in addition to femtoscopic
correlations.  In these cases,
the most common techniques form a background from unlike-signed pairs, with
resonance regions excluded with cuts~\cite{Abreu:1992gj} or normalized with a
correlation of like- to unlike-signed pairs from a Monte
Carlo~\cite{Abbiendi:2000gc}. Other experiments have constructed a
background using only Monte Carlo generated pairs~~\cite{UribeDuque:1993gz}.  A few
experiments have investigated backgrounds formed by swapping~\cite{Avery:1985qb} or
reversing momentum components relative to a jet-axis~\cite{Abreu:1992gj}, but these
methods are not widely used.  For detectors with symmetrical acceptance, such as the
STAR TPC~\cite{Anderson:2003ur}, momentum conservation effects may be eliminated by
mixing pairs from the same event, with the lab momentum of one particle flipped~\cite{Stavinskiy:2004xx}.
Backgrounds constructed from single-particle
distributions as formally defined by 
Equation~\ref{eq:FormalDefinition} have been used for heavy ion collisions at lower
energies and shown to be consistent with the more commonly used event-mixing
technique~\cite{Lisa:1991xx}.

In order to avoid inducing artificial structure in the correlation function, the particles forming
pairs in the background distribution should originate from parent events with the same event characteristics.
The parent events should have similar vertex positions to within the experimental
resolution.
Because detector acceptances can vary with time (e.g., components may fail for some runs),
parent events should have been measured close in time to each other; this is usually
easiest in any case, because event mixing is done on the fly as time-ordered data
is read sequentially.

Parent events whose particles are mixed should also have the same single-particle momentum distributions.
Thus, they should have similar centralities and orientations of the reaction plane.  For example, mixing
particles from
events with very different $p_T$ slopes or directions of preferred emission (elliptic flow) would produce
differences between $A^{ab}({\bf q})$ and $B^{ab}({\bf q})$ even in the absence of physical correlations.
because almost all analyses to date have ignored these potential biases, it is comforting that
they make little difference in practice~\cite{Adams:2004yc}.

The list of event classes given here is by no means exhaustive.  One
can expect future analyses to incorporate the orientation of high
$p_T$ particles (jet axis) or any other event-related observable.

The procedure for deciding how many events to mix remains something of an art and
involves optimizing over the range of data runs, bin width, and statistics.  In
order to minimize statistical errors, one typically forms approximately ten times
the number of pairs in the background as in the signal.  For the special case when
all possible combinations are formed, the variance of a particular relative momentum
bin is proportional to $n^\frac{3}{4}$, where $n$ is the number of entries in the
bin~\cite{Zajc:1984vb}.  However, as the number of pairs formed is reduced, the
variance per bin approaches the $n^\frac{1}{2}$ value expected for Poisson
statistics~\cite{Ahle:2002mi}.  It is possible that non-Poisson fluctuations persist
in the co-variance between different bins, but this has not yet been investigated.

Once the pairs have been mixed, the background must be subject to the same
2-track cuts that have been applied to the signal.  For example, the
exact same track merging cuts or minimum separation on a detector must
be applied to both signal and background.

\subsection{Corrections}

Corrections to the correlation function fall into three categories: finite
resolution effects, mis-identified particle contamination, and compensation for
deficiencies in the background.  

The first category concerns 
single-track momentum resolution, and reaction-plane resolution.  We
consider finite momentum resolution corrections first.  Typically, momentum
resolutions are on the order of 1\%.  One approach is to correct for momentum
resolution by a double ratio of the ideal correlation function generated from a
Monte Carlo simulation with perfect momentum resolution divided by a Monte
Carlo correlation function with momentum resolution turned on.  The femtoscopic
weights are inserted into the simulations iteratively until the fitted radii
converge~\cite{Boggild:1993zj,Lisa:2005vw,Lisa:2000no,Adams:2004yc}.  A second approach is similar, but
corrects only the Coulomb interaction term, which is most greatly affected by 
momentum resolution effects~\cite{Adler:2004rq}.  In both cases, the
corrections change the fitted radii by only $<$5\%.

As discussed in Section~\ref{sec:systematics-phi}, azimuthally sensitive
analyses~\cite{Lisa:2000xj,Wells:2002phd,Adams:2003ra} measure
oscillations in correlations as a function of emission angle with respect to the reaction plane.
Finite resolution effects in the reaction plane angle~\cite{Poskanzer:1998yz} artificially
reduce the oscillation strengths.  Methods have been developed to correct the distributions
$A^{ab}({\bf q})$ and $B^{ab}({\bf q})$ for these effects~\cite{Heinz:2002au,Borghini:2004ra}.

Another type of correction accounts for the inclusion of mis-identified and secondary-particle contamination.
For example, electrons may be mistakenly identified as $\pi^-$ mesons.  It is usually assumed that the
mis-identified or secondary particles are uncorrelated with other particles, so the net effect on the
correlation function is to damp all structure uniformly in ${\bf q}$.
For purely Gaussian correlations (see Section~\ref{sec:Fitting}),
this effect is absorbed wholly into the $\lambda$ factor discussed in Section~\ref{sec:lambda};
homogeneity lengths---derived from the width of the Gaussian correlation---are unaffected by the reduction
in its strength.
In many cases, however, the homogeneity length is extracted largely from the strength of the correlation, and so
contamination effects must be removed.  In the general case, for which the purity $\rho$ depends on the relative
momentum, the correlation function is corrected according to
$C^{\rm true}({\bf q}) = (C^{\rm raw}({\bf q})-1)/\rho({\bf q}) + 1$~\cite{Adams:2003qa,Adams:2004yc}.

It is more difficult to correct for correlated contamination.  For example, if cuts cannot completely distinguish
primary protons from those coming from $\Lambda$ decay, then measured $p-\Lambda$ correlations will contain contributions
from $\Lambda-\Lambda$ correlations.  Unlike the white-noise contamination discussed above, this introduces structure
into $C^{p\Lambda}({\bf q})$ that can be accounted for only with detailed simulations.  Such corrections will become
more important at RHIC due to copious resonance production, and especially for baryon correlation measurements, in which
the heavy daughter carries most of the momentum of the parent resonance.

The last category of corrections are applied to fix deficiencies in the background
distribution.  
This includes corrections to account for two-particle inefficiencies, which
have been discussed in the previous Section.
  A second correction of this type deals with the residual
signal correlation that is present in all backgrounds derived from events that
contribute to the signal.  The residual correlation arises because femtoscopic
correlations can modify the single particle distributions.  This is especially true
for small-aperture spectrometers.  This effect can be removed with an iterative
procedure~\cite{Zajc:1984vb,Bearden:1998aq}, however, for many large experiments
the induced error is often 1\% or less, and it is easier to fold this into the
systematic errors~~\cite{Adler:2004rq}.

\subsection{Fitting}
\label{sec:Fitting}

After the application of all cuts and corrections, the correlation is formed
according to Equation~\ref{eq:ExperimentalDefinition} and then fit to determine spatial
parameters.  As described in Section~\ref{sec:theoryBasics}, there are three
approaches to fitting the correlation function: fitting to a simplified
Gaussian form with strong and Coulomb interactions neglected or factored out,
fitting to a convolution of the full kernel convoluted to a parameterization of
${\cal S}({\bf r})$, and inverting the kernel to fit a source image.  The
simplified Gaussian fits based on Equation~\ref{eq:C2Gauss} are limited to
correlations of identical pions, kaons, and photons, but it has been the most
widely used method to date because of the computational demands of the other
methods.  We expect its use to continue for the large systematic studies in
which binning in centrality, reaction plane, and $k_T$ leads to fits of more than one hundred
separate correlation functions for a single colliding system.

However, the functional form of the Gaussian parameterization used by
experimentalists has evolved over the years.  Before reviewing the most recent
functional forms, it is necessary to review the treatment of the Coulomb
interaction and the fraction of pairs coming from the source that contribute
to the femtoscopic correlations.  Both were first introduced into the
literature by Zajc~\cite{Zajc:1984vb} in the form of the Gamow factor given in
Section~\ref{sec:theoryBasics} and an empirical parameter $\lambda$ to account
for the observation that not all pairs exhibit femtoscopic correlations.  With
steady improvements in data quality and CPU speed, the Gamow factor has been
replaced with a calculation of an squared unsymmetrized Coulomb wave for a
finite Gaussian source.  The improvements in data quality have also led to a
self-consistent treatment of $\lambda$ with respect to both Coulomb and
Gaussian components of the fit function~\cite{Bowler:1991vx,Sinyukov:1998fc}.
For this to be accurate, we must assume that the source is fully chaotic, an
assumption that has recently been verified with three-pion
correlations~\cite{Adams:2003vd,Heinz:1997mr}.  The non-femtoscopic pairs
consist of mis-identified particles and particles that emanate from too far
from the source for the correlation to be resolved experimentally.  The region
far from the source has been referred to the source halo, to differentiate it
from the core.  The correlation fit function is therefore given by
Equation~\ref{eq:C2CoreHalo},
\begin{equation}
C({\bf q}) = N \left[ \lambda G({\bf q}) F({\bf q}) + 
	\left( 1 - \lambda \right) \right] ,
\label{eq:C2CoreHalo}
\end{equation}
where $N$ is the overall normalization, $F$ is the Coulomb component, and $G$
is the Gaussian form for the un-damped correlation function,
Equation~\ref{eq:C2Gauss} for out-long-side coordinates, or
Equation~\ref{eq:yk} for Yano-Koonin variables.

\begin{SCfigure}[0.6][t!]
\includegraphics[width=0.6\textwidth]{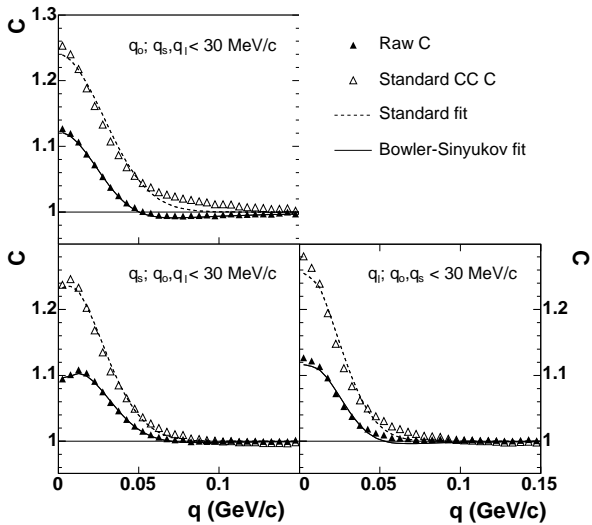}
\caption{\label{fig:c2} Projections of a three-dimensional correlation function
  (integrated over 0-30~MeV/c in orthogonal components) for low-$k_T$ $\pi^-$
  pairs for 200~GeV central Au+Au collisions~\cite{Adams:2004yc} (filled
  symbols) with fit function.  Open symbols include correction for Coulomb
  interaction among all pairs.  Projections were generated according to the prescription described in~\cite{Chacon:1991ri}.}
\end{SCfigure}

Figure~\ref{fig:c2} shows projections of a $\pi^--\pi^-$ correlation function
measured by the STAR collaboration~\cite{Adams:2004yc}.  The filled symbols are
the measured correlation function corrected for momentum resolution only and
fit with Equation~\ref{eq:C2CoreHalo}.  The open symbols have been
overcorrected by applying to all pairs the Coulomb correction for the fitted
source dimensions.  Depending on the shape of the correlation and degree of
experimental contamination, extracted homogeneity lengths may vary by up to
$\sim 15\%$ if the correlation function is overcorrected.
%

For proton-proton correlations and non-identical particle correlations, direct
fits are performed by convoluting the full kernel with a parameterized source.
For these analyses, the paucity of statistics has been more of a limitation
than the relatively modest demands in CPU power.  The examples given in
Section~\ref{sec:systematics} are all for one-dimensional analyses, but recent
data from RHIC will soon be analyzed in multi-dimensions.

The ability to image the source by inverting the kernel is a relatively recent
development, but one with very general applications.  Because the source is
parameterized by a series of B-splines, it is a very general form which is
sensitive to non-Gaussian shapes.  To date, source imaging has been performed
only with one-dimensional correlations, but like with the direct fits, a
multi-dimensional kernel will soon be possible \cite{Brown:2004bh,Danielewicz:2005qh}.

Non-Gaussian effects were reported in the first pion correlation measurement at RHIC~\cite{Adler:2001zd}.
With higher statistics, the STAR Collaboration has studied the issue in greater detail~\cite{Adams:2004yc}, performing
a functional expansion (the so-called Edgeworth expansion~\cite{Csorgo:2003uv}) about a Gaussian shape.
Although significant non-Gaussian contributions were reported, the dominant length scales were already extracted
in the purely Gaussian fits.

\subsubsection{Minimization}

A simple chi-squared test is inappropriate for fitting correlation functions
because the ratio of two Poisson distributions is not itself Poisson
distributed, especially when taking the ratio of small numbers.  For this reason, a
log-likelihood fit function of the form given in Equation~\ref{eq:pml2} is preferred.  
\begin{equation}
\label{eq:pml2}
\chisq_{PML} = -2 \left[ 
A \ln \left( \frac{ C \left( A+B \right)}{A \left( C+1 \right)}\right)
+ B \ln \left( \frac{A+B}{B \left( C+1 \right)} \right) \right],
\end{equation}
where {\it A}, {\it B}, and {\it C} were introduced in Equation 35.
This equation derived from the principle of maximum likelihood assuming that
both signal and background are Poisson distributed~\cite{Ahle:2002mi}.  The full
derivation of Equation~\ref{eq:pml2} and comparison to earlier log-likelihood functions is
given in~\cite{Ahle:2002mi}.


%

\section{Measured Femtoscopic Systematics}
\label{sec:systematics}

The first systematic study to compare femtoscopic measurements across several
systems and experiments was performed almost 20 years ago with data from
intermediate-energy heavy ion collisions at the Bevalac~\cite{Bartke:1986mj}.  The data, taken
 from experiments with different acceptances, triggers, and analysis techniques,
were sufficient to demonstrate a crude $A^{1/3}$ scaling of the one-dimensional radii,
indicating that spatial scales were indeed being probed.  The first femtoscopic
measurements for relativistic heavy ion collisions were presented by the NA35
Collaboration at the Quark Matter meeting in
Nordkirchen~\cite{Humanic:1988ny,Bamberger:1988kd}.  More detailed
measurements followed with the availability of sulphur and silicon beams at the
SPS~\cite{Boggild:1993zj} and AGS~\cite{Abbott:1992rt,Barrette:1994pi}.  

Since then, increasingly sophisticated experiments at the AGS, SPS, and RHIC have
performed femtoscopic measurements corresponding to a wide range of control
parameters.  The experimental community performing the measurements has reached
critical mass and matured substantially; a common language and knowledge base has
developed concerning sometimes subtle details in performing and interpreting
femtoscopic measurements.  The result of this effort is a striking degree of
consistency across experiments in regions of phase space where acceptances overlap
and meaningful generation of systematics across experiments.  Large-statistics data
sets routinely allow three-dimensional correlation measurements with small
statistical error bars.  Systematic errors, which now dominate the experimental
errors, have been reduced to the level of $\sim 5\%$, or $\sim 0.25$~fm for most
measurements.  It is no exaggeration to state that femtoscopic measurements have
become a precision tool.

Here, we cover the most important systematics of femtoscopic measurements from the
AGS, SPS, and RHIC.  We discuss only generally the physics probed by a given
systematic, appealing to intuitive schematic models such as the
blast wave~\cite{Retiere:2003kf}.  Full interpretations and comparisons to dynamic
models are given in Section~\ref{sec:interpretations}



\subsection{System size: $N_{\rm part}$ and Multiplicity}
\label{sec:systematics-Npart}

As discussed earlier, femtoscopic radii probe homogeneity regions, and not the
entire source (hereafter, the term {\it source} will be used to refer to the entire 
source of particle emission).  Nevertheless, the 
claim that two-particle correlations probe spatial scales would be given little
credence if the radii did not exhibit a strong, positive correlation with system
size.  Therefore, measuring the systematic variation of the radii vs. system
composition and centrality represents the most basic test of both theoretical and
experimental femtoscopic techniques.


\begin{figure}[t!]
\centerline{\includegraphics[width=0.7\textwidth]{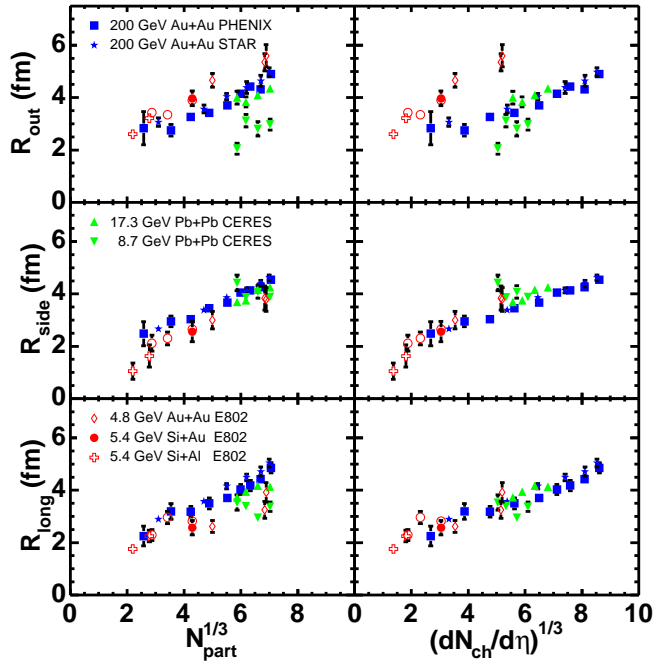}}
\caption{\label{fig:HBTcent}
Pion source radius dependence on number of participants ({\it left}) and on charged particle multiplicity ({\it right}).
Data are for for Au+Au (Pb+Pb) collisions at several values of \roots, and also for Si+A collisions at the
lowest energy.
Average transverse momentum $\langle k_T \rangle \sim 450$~MeV/c for the PHENIX data and $\sim 390$~MeV/c
for the others.  Data from~\cite{Ahle:2002mi,Adamova:2002wi,Adler:2004rq,Adams:2004yc}.}
\end{figure}

Coalescence studies~\cite{Barrette:1994tw} and two-proton measurements at the
AGS~\cite{Barrette:1999qn} and SPS~\cite{Boggild:1998dx} unambiguously
demonstrate that nucleon homogeneity lengths increase with decreasing impact
parameter and/or increasing projectile mass, continuing the trend mapped at
lower energies~\cite{Lisa:1993xh,Kotte:2004yv}, where directional cuts have
allowed measurement of the shape of the homogeneity
region~\cite{Lisa:1993xx,Lisa:1994xx,Kotte:1997kz}.  More detailed information
comes from pion correlations at relativistic energies, for which
three-dimensional analyses allow partial isolation of purely geometrical
effects.  The centrality dependence of Bertsch-Pratt source radii are shown in
Figure~\ref{fig:HBTcent} for a wide range of collision energies.  The left
panels show the dependence on the number of participating nucleons, \Npart, a
generalization of the $A^{1/3}$ linear scaling of nuclear radii used to
approximate the initial overlap geometry.  All of the radii exhibit a linear
scaling in \Npthree, most with finite intercepts. Only the slope of
the \Rlong~dependence shows a significant increase from the AGS to RHIC,
consistent with a lifetime that increases with both centrality and
$\sqrt{s_{NN}}$.  The trend of increasing \Rlong~with increasing \roots~is
reversed for $\roots < 5$~GeV~\cite{Lisa:2000no}.

The right panels of Figure~\ref{fig:HBTcent} show the same radii as a function of
\dNthree.  The primary motivation for exploring the \dNthree dependence is its
relation to the final state geometry through the density at freeze-out.  However,
the two scaling quantities are highly correlated.  In fact, the values of $dN_{ch}/d\eta$ shown on the
right side of Figure~\ref{fig:HBTcent} were derived from $N_{\rm part}$ using the
$N_{\rm part}^{\alpha}$ parameterizations given in~\cite{Adler:2004zn}, and conversely,
the \Npart~values are often calculated from multiplicity distributions using a
Glauber model.  Given this caveat, the \Rside~and \Rlong values exhibit a 
linear dependence on \dNthree, again with finite intercepts.  The strong uniformity
from \roots~of 5 to 200 GeV leads one to believe that the approximate $N_{\rm part}$ scaling
(initial overlap geometry) is a result of the scaling with multiplicity (final
freeze-out geometry) and not the other way around.  

The parameter \Rout, which mixes spatial and temporal information (see Section 2.5), increases with multiplicity at each given collision energy, but does not follow a universal curve. However, the strikingly \roots-independent multiplicity scaling of the geometric radii \Rside~and \Rlong strongly suggests that the observed increase of these radii with collision energy for \roots~$>$ 5 GeV (see Section 5.1) is due simply to the rise of multiplicity with collision energy. This trend, as well as its violation at \roots~$<$ 5 GeV, has been interpreted in terms of changing chemical composition of the source as the system evolves with energy from baryon to meson dominance~\cite{Adamova:2002ff}.              

We note that the systematics in system size represent an initial sanity check
for the femtoscopic technique.  The obvious direct connection of the radii to
the source geometry estimated in two ways refutes suggestions~\cite{Csorgo:1995bi} that
smaller, non-geometric length scales dominate experimentally extracted transverse
radii.


\subsection{Source Shape: Pair emission angle relative to $\hat{b}$}
\label{sec:systematics-phi}

The variation of femtoscopic radii with the pair emission angle relative to
$\vec{b}$ (\phipair) can be used to probe the three-dimensional shape of the
source~\cite{Voloshin:1995mc,Voloshin:1996ch,Wiedemann:1997cr,Heiselberg:1998ik,Heiselberg:1998es,Lisa:2000ip,Heinz:2002au,Retiere:2003kf}.
The anisotropic shape transverse to the beam direction---the
coordinate-space analog to the elliptic flow characterizing momentum-space---gives rise to $\cos(n\phipair)$ ($n$ even) oscillations in the squared transverse source radii \RoSq, \RsSq, \RosSq~\cite{Wiedemann:1997cr,Heinz:2002au}.

\begin{figure}[t!]
\begin{minipage}[t]{0.48\textwidth}
\centerline{\includegraphics[width=1.0\textwidth]{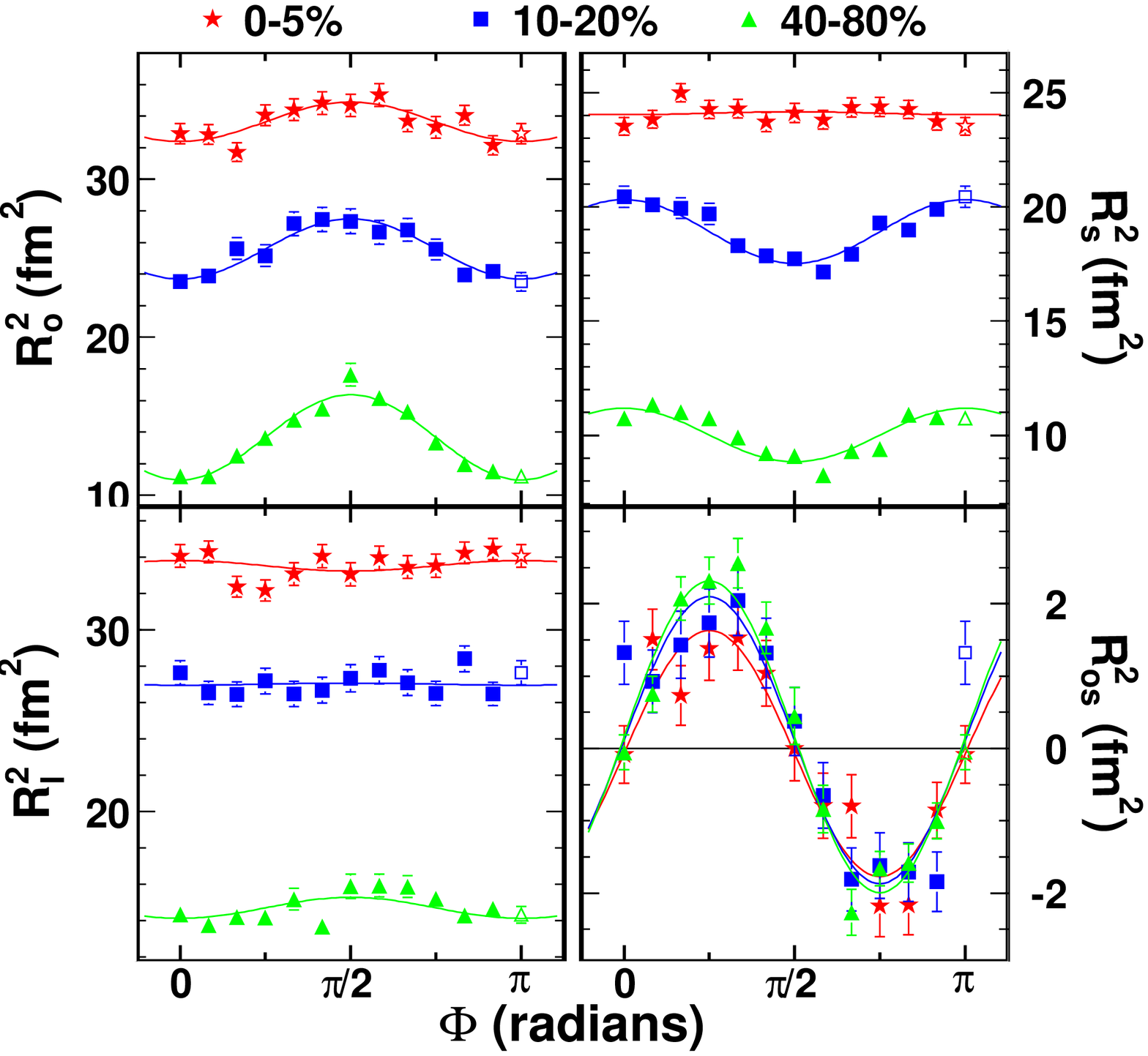}}
    \caption{Squared source radii measured at mid-rapidity in Au+Au collisions
      at RHIC, as a function of pair emission angle relative to the reaction
      plane.  Data for three centralities are shown.  Figure
      from~\cite{Adams:2003ra}.  }
    \label{fig:STARasHBTfig1}
\end{minipage}
\hspace{\fill}
\begin{minipage}[t]{0.49\textwidth}
\centerline{\includegraphics[width=1.0\textwidth]{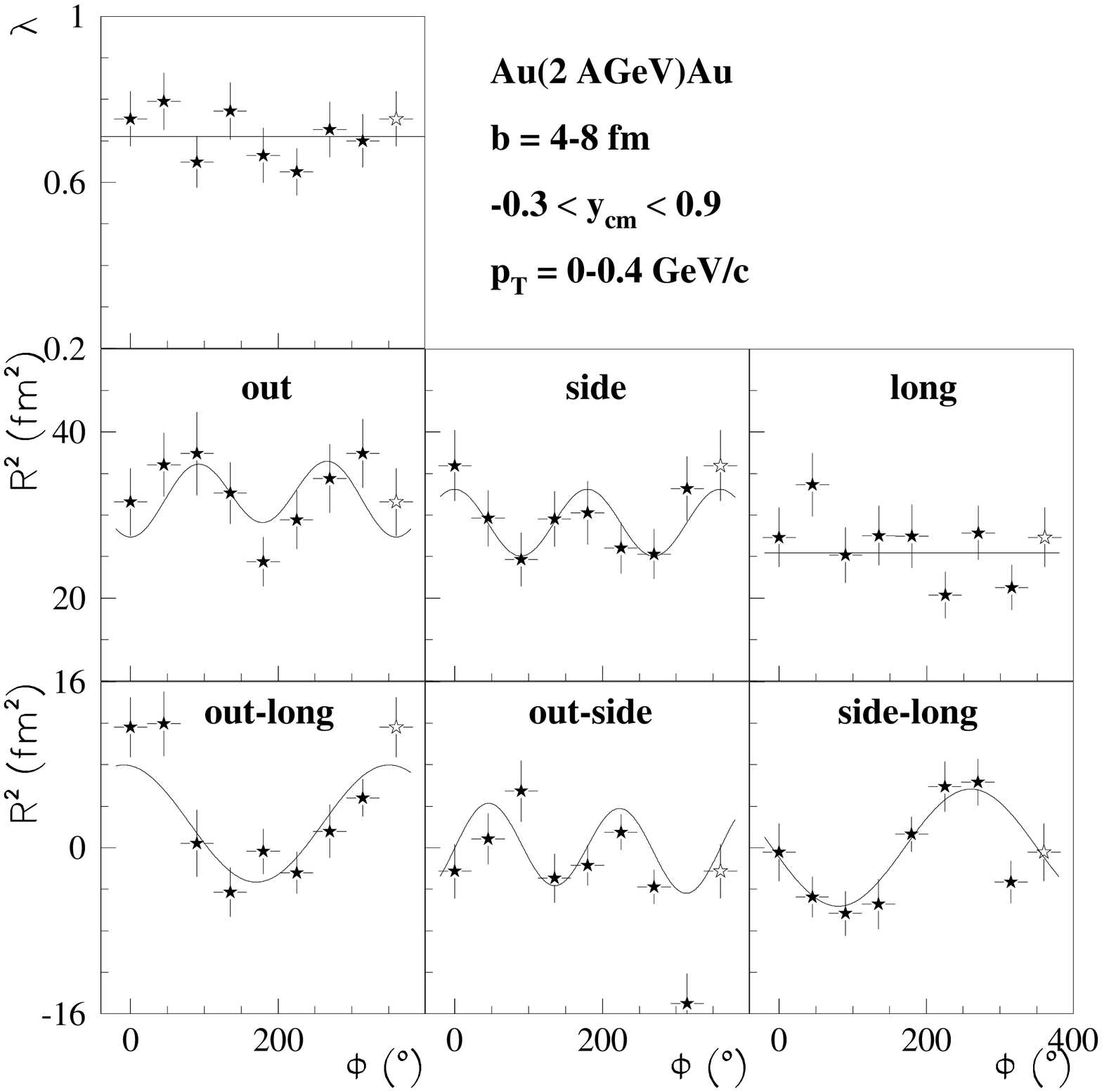}}
    \caption{Squared source radii measured near mid-rapidity in mid-central
      Au+Au collisions at lowest-energy AGS settings, as a function of pair
      emission angle relative to the reaction plane.  Figure
      from~\cite{Lisa:2000xj}.  }
    \label{fig:E8952GeVasHBT}
\end{minipage}
\end{figure}

Just as one expects the source (and homogeneity regions) to be larger 
for decreasing $|\vec{b}|$, one also expects it to be rounder,
reflected by small oscillations of the radii.  Figure~\ref{fig:STARasHBTfig1} for mid-rapidity pions 
from Au+Au collisions at RHIC confirms this expectation.  As $|\vec{b}|$ increases, the oscillations
indicate a transverse source increasingly elongated out of the reaction plane~\cite{Adams:2003ra}.\footnote{The out-of-plane nature of
the elongation may be read directly from Figure~\ref{fig:STARasHBTfig1}.  Ignoring collective flow or opacity effects~\citep[e.g.][]{Lisa:2000ip}
an out-of-plane-extended Source would produce $\RsSq(\phipair=0^\circ) > \RsSq(\phipair=90^\circ)$, as seen in Figure~\ref{fig:STARasHBTfig1}.
Collective flow effects complicate this picture~\cite{Wiedemann:1997cr,Retiere:2003kf}, but the sign of the oscillations
are determined by geometric, not dynamic, effects for realistic sources at RHIC~\cite{Heinz:2002sq,Retiere:2003kf}.}

The strong in-plane expansion~\cite{Ackermann:2000tr} does not fully convert the initial out-of-plane (overlap) geometry
into an in-plane-extended source at freeze-out. This suggests a rather short evolution time; in essence, the system did not have time
to reverse its deformation.  However, this is only a hint, and a full dynamical transport calculation is required to extract physical
timescales~\cite{Heinz:2002sq}.

Whereas at the highest RHIC energy, the freeze-out anisotropy is $\sim \frac{1}{3}$ of the initial~\cite{Adams:2003ra},
at low AGS energies, the final anisotropy is consistent with that of the initial overlap region~\cite{Lisa:2000xj}, or perhaps
slightly lower.  because elliptic flow vanishes---changes sign---at these energies 
\cite{Pinkenburg:1999ya},
these trends make intuitive sense and suggest an underlying connection to the evolution dynamics.
It would be desirable to map the source anisotropy at intermediate (AGS and SPS) energies, for which 
there may be interesting changes in the space-time systematics.  
At these energies, there have been
intriguing hints of asymmetries in the homogeneity regions 
for pions~\cite{Miskowiec:1995df,Filimonov:1999ya,Nishimura:1999wz,Aggarwal:2000uj}
and protons~\cite{Panitkin:1999yd}, and in the proton-pion
separation~\cite{Filimonov:1999ya,Miskowiec:1998ms}, although they have not been finalized.

%

If the impact parameter direction $\hat{b}$---not simply the $2^{\rm nd}$-order event-plane angle (unambiguous only over
a range $[0,\pi]$)---is known, then more detailed information
may be obtained.  In the left panel of Figure~\ref{fig:CrossTerm}, the source is tilted with respect to the beam axis, toward $\hat{b}$.
 Just as anisotropic azimuthal geometry in the transverse plane is related to the structure of elliptic flow~\cite{Heinz:2002sq,Kolb:2003dz},
a tilted geometry can reveal important information on the underlying nature of directed flow~\cite{Csernai:1999nf,Brachmann:1999xt,Lisa:2000ip,Lisa:2000xj}.
The structure in $R^2_{o,l}$ and $R^2_{s,l}$ shown in Figure~\ref{fig:E8952GeVasHBT} is generated by this tilt.
The spatial tilt has been measured only at low AGS energies~\cite{Lisa:2000xj}; a measurement at RHIC might reveal exotic geometric
configurations generated by quark gluon plasma (QGP) formation~\cite{Brachmann:1999xt,Magas:2000cm}   
 and
would impact the important issue of boost-invariance at mid-rapidity.

\subsection{Boost invariance : $Y_{\pi\pi}$}
\label{sec:systRapidity}

In high-energy hadronic collisions, the initial parton distribution is expected
to be approximately flat in rapidity. This
momentum rapidity distribution may correspond to producing matter that initially
exhibits a boost-invariant Hubble-type scaling correlation between longitudinal
flow velocity and space-time points, $v_{L,flow} =
z/t$~\cite{Shuryak:1980tp,Bjorken:1982qr}. Relativistic hydrodynamics preserves
boost-invariance of the initial conditions throughout its dynamical
evolution~\cite{Bjorken:1982qr}. The combination of the above arguments
underlies expectations that in ultra-relativistic heavy ion collisions,
particle production emerges from boost-invariant longitudinal flow, and that
$dN/dy$ exhibits an approximately boost-invariant plateau around
mid-rapidity. However, an extended plateau has never been observed from AGS
through RHIC energies~\cite{Back:2004je}.

because correlation measurements access spatio-temporal information, the question
arises~\cite{Wu:1996wk,Heinz:1996rw} whether they allow us to test the relation
$v_{L,flow} = z/t$ between the space-time rapidity and the momentum rapidity of
the source. For boost-invariant sources, one can show that the pair momentum
rapidity is equal to the Yano-Koonin source velocity, which is directly obtained
from the Gaussian radius parameters, $Y_{\pi\pi} \approx Y_{L,flow} \equiv
\tanh^{-1} v_{L,flow}$.  However, even if the source density distribution shows
significant deviations from boost-invariance, this relation still holds
approximately as long as the velocity profile is boost invariant, and $k_T$ is
sufficiently large~\cite{Wu:1996wk}.

\begin{SCfigure}[0.49][t!]
{\includegraphics[width=0.6\textwidth]{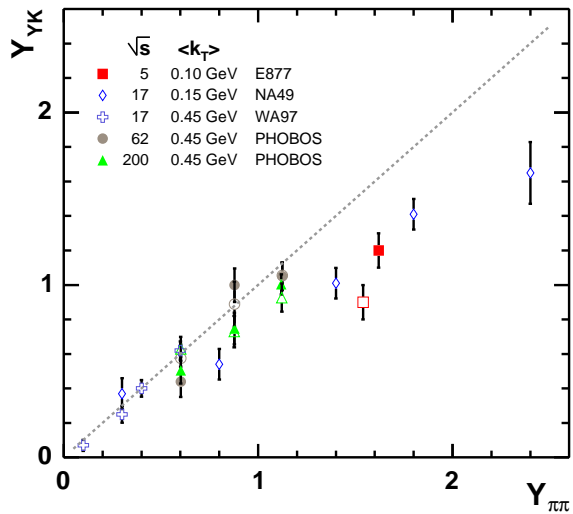}}
\caption{\label{fig:HBTydep} The Yano-Koonin source rapidity is shown as a
  function of the pion pair rapidity for central Au(Pb)+Au(Pb) collisions over
  a broad range of energies (open symbols are for $\pi^-\pi^-$, closed symbols
  for $\pi^+\pi^+$).  Both quantities are in the center-of-mass frame
  of the colliding system.  Data
  from~\cite{Miskowiec:1996xa,Appelshauser:1997rr,Antinori:2001yi,Back:2004ug}.
}
\end{SCfigure}

Figure~\ref{fig:HBTydep} reveals a roughly universal dependence of \YYK~ on $Y_{\pi\pi}$
for pions from central collisions, depending weakly, if at all, on $\sqrt{s_{NN}}$ 
(see Section~\ref{sec:CrossTerms}).  This trend is 
particularly striking given the very different center-of-mass projectile rapidities ($\sim 1.55$ 
and 5.5 for $\sqrt{s_{NN}}=5$ and 200~GeV, respectively) and corresponding
widths of the pion distributions $dN/dy$.  

By way of caution, we note that the results from RHIC are limited to the region $Y_{\pi\pi}<1.2$, and that the 
deviations from boost-invariance are mostly in the lower energy data.  Extending the RHIC 
results to more forward rapidities would provide a important test for both the velocity 
scaling at RHIC and the energy-independence that is exhibited in Figure~\ref{fig:HBTydep}.

For central collisions, the roughly universal behavior approximately 
obeys the boost-invariant consistency relationship discussed above. Moreover, $Y_{L,flow}$ shows
a significant $k_T$ dependence and falls below the linear relation in particular for small $k_T$
~\cite{Antinori:2001yi}. Qualitatively, this is consistent with blast-wave models in which a 
boost-invariant longitudinal flow is superimposed on a source density distribution of
finite longitudinal width. However, a full dynamical understanding of the dependence is
missing so far.  The flat \YYK~ dependence on $Y_{\pi\pi}$ measured at
the SPS~\cite{Antinori:2001yi} for the most peripheral collisions is
counter-intuitive, and requires further study.

The question of whether the source has boost-invariant space-time structure is
an important one.  There are many reports of very short evolution timescales
(lifetimes) based on fits to the data with Equation~\ref{eq:meantauestimate},
which is based upon an assumption of boost-invariance~\cite{Makhlin:1987gm}.
Relaxation of that assumption might lead to considerably larger
estimates~\cite{Renk:2003gn}.

\subsection{Collective dynamics: $k_T$ and particle mass}
\label{sec:pTmass}

As discussed in Section~\ref{sec:transverseFlow}, the dynamic substructure of the source is encoded in
space-momentum (\xp) correlations.  Longitudinal \xp ~correlations, encoded in $\Rlong(k_T)$, are
generally acknowledged~\cite{Makhlin:1987gm,Retiere:2003kf} to reflect longitudinal flow.
because all transverse correlations are generated in the collision itself,
considerably more attention has generally been paid to the transverse substructure
than to the longitudinal flow discussed in Section~\ref{sec:systRapidity}.  

The most common explanation for transverse \xp ~correlations is collective transverse flow~\cite{Pratt:1984su}.
These correlations have mostly been
studied through pion correlations, but transverse flow implies also a systematic trend as the particle mass is varied.

\subsubsection{$k_T$ dependence of pion radii}

\begin{figure}[t!]
\centerline{\includegraphics[width=1.0\textwidth]{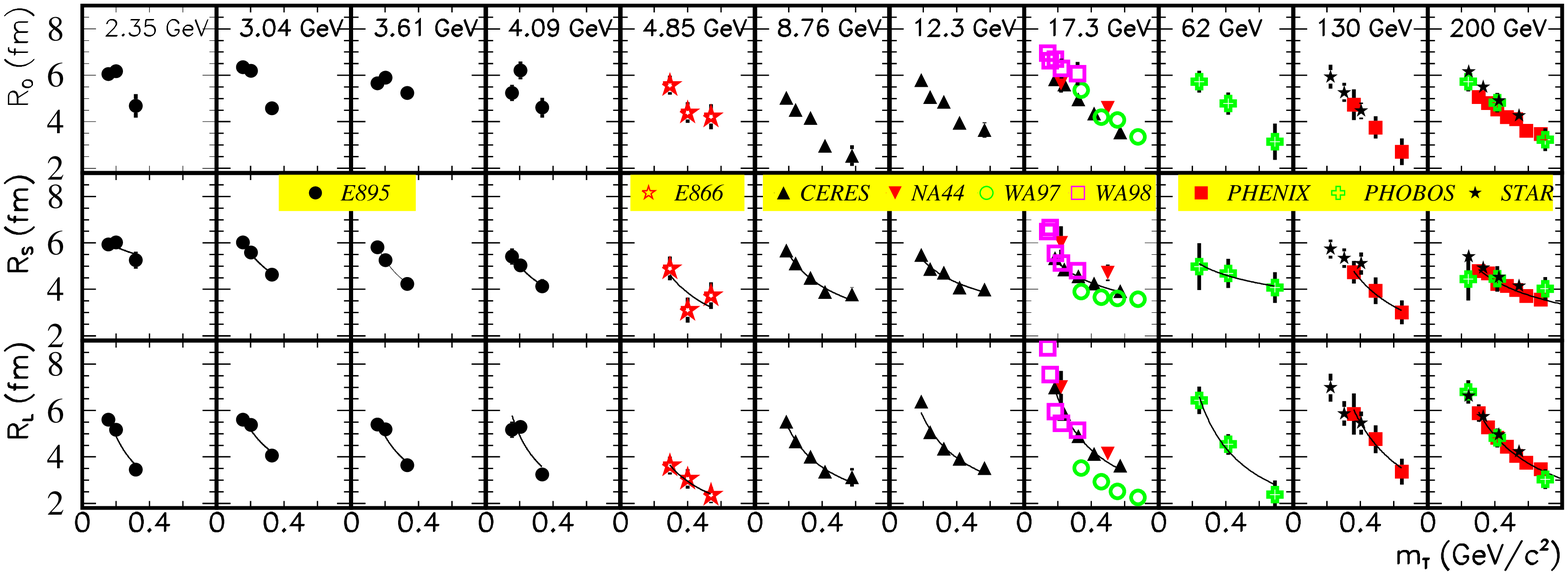}}
\caption{\label{fig:kTdepAllEnergies}
World data set of published $m_T$ dependence of pion Bertsch-Pratt radii near mid-rapidity from Au+Au (Pb+Pb) collisions.
Centrality selection is roughly top 10\% of cross-section, but varies somewhat with experiment.
Data from~\cite{Lisa:2000no,Ahle:2002mi,Adamova:2002wi,Bearden:1998aq,Antinori:2001yi,Aggarwal:2002tm,Adcox:2002uc,Adler:2004rq,Back:2004ug,Adler:2001zd,Adams:2004yc}
Lines represent parameterized fits; see text for details.  
}
\end{figure}

Collective flow generates a characteristic fall-off of the pion source radii
with $k_T$, which is ubiquitously observed in data.  Final results for the
$k_T$-dependence of Gaussian radii from central Au+Au (Pb+Pb) collisions exist
at the AGS~\cite{Lisa:2000no,Ahle:2002mi},
SPS~\cite{Appelshauser:1997rr,Bearden:1998aq,Aggarwal:2002tm,Antinori:2001yi,Adamova:2002wi},
and
RHIC~\cite{Adler:2001zd,Adcox:2002uc,Adams:2003ra,Adams:2004yc,Back:2004ug,Adler:2004rq}.
As is clear from Figure~\ref{fig:kTdepAllEnergies}, aside from a small variation in overall scale (discussed later), 
the $k_T$ dependence is startlingly similar for all energies.

\begin{figure}[t!]
\begin{minipage}[t]{0.49\textwidth}
 \centerline{\includegraphics[width=1.0\textwidth]{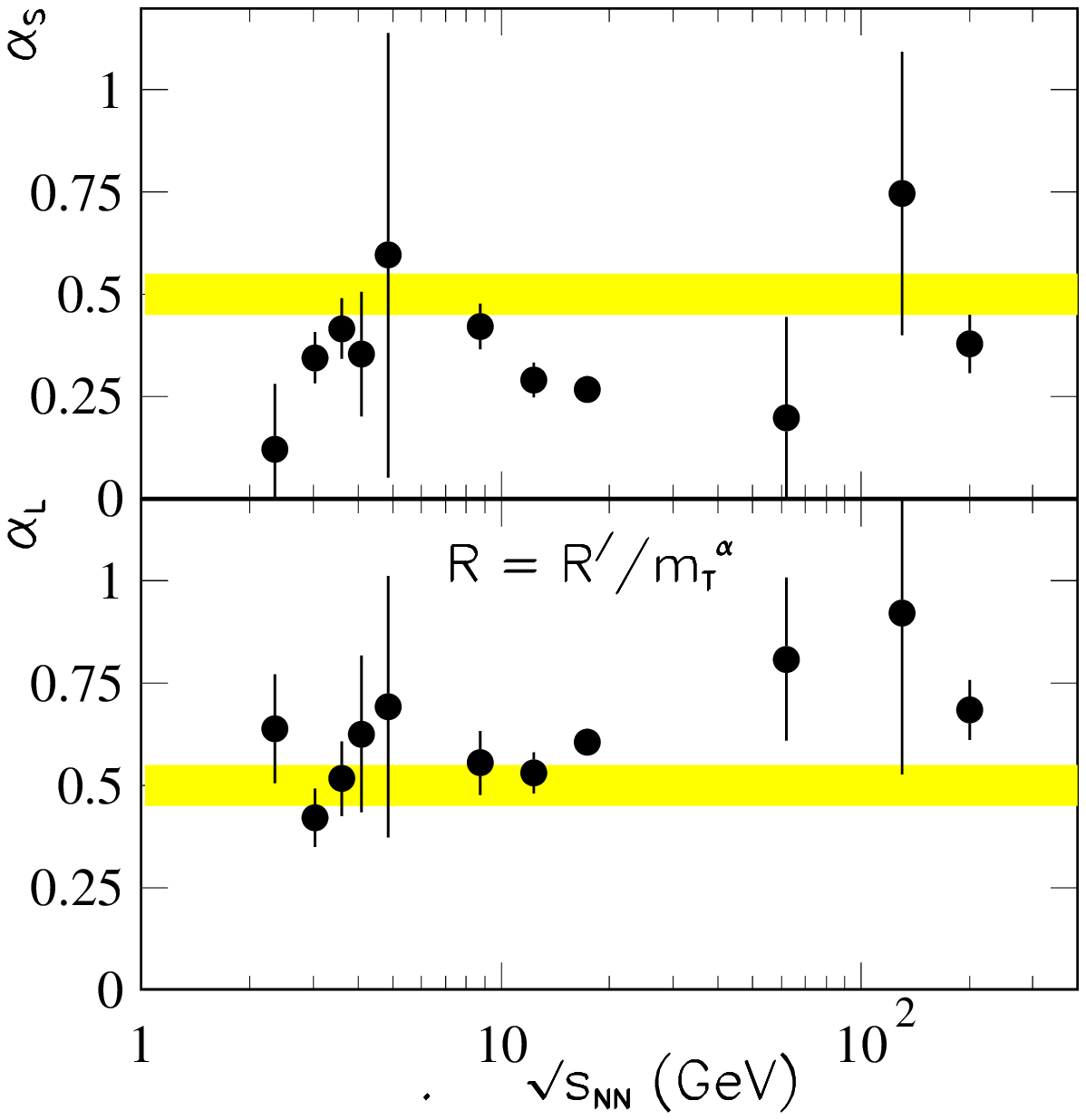}}
 \caption{\label{fig:PowerLawFits}
 The excitation function of $m_T$ power-law fall-off of pion source radii ($R_i(m_T) \sim m_T^{-\alpha_i}$).  
 Shaded regions show $\alpha_i = 0.5 \pm 0.05$.
 }
\end{minipage}
\hspace{\fill}
\begin{minipage}[t]{0.49\textwidth}
 \centerline{\includegraphics[width=1.0\textwidth]{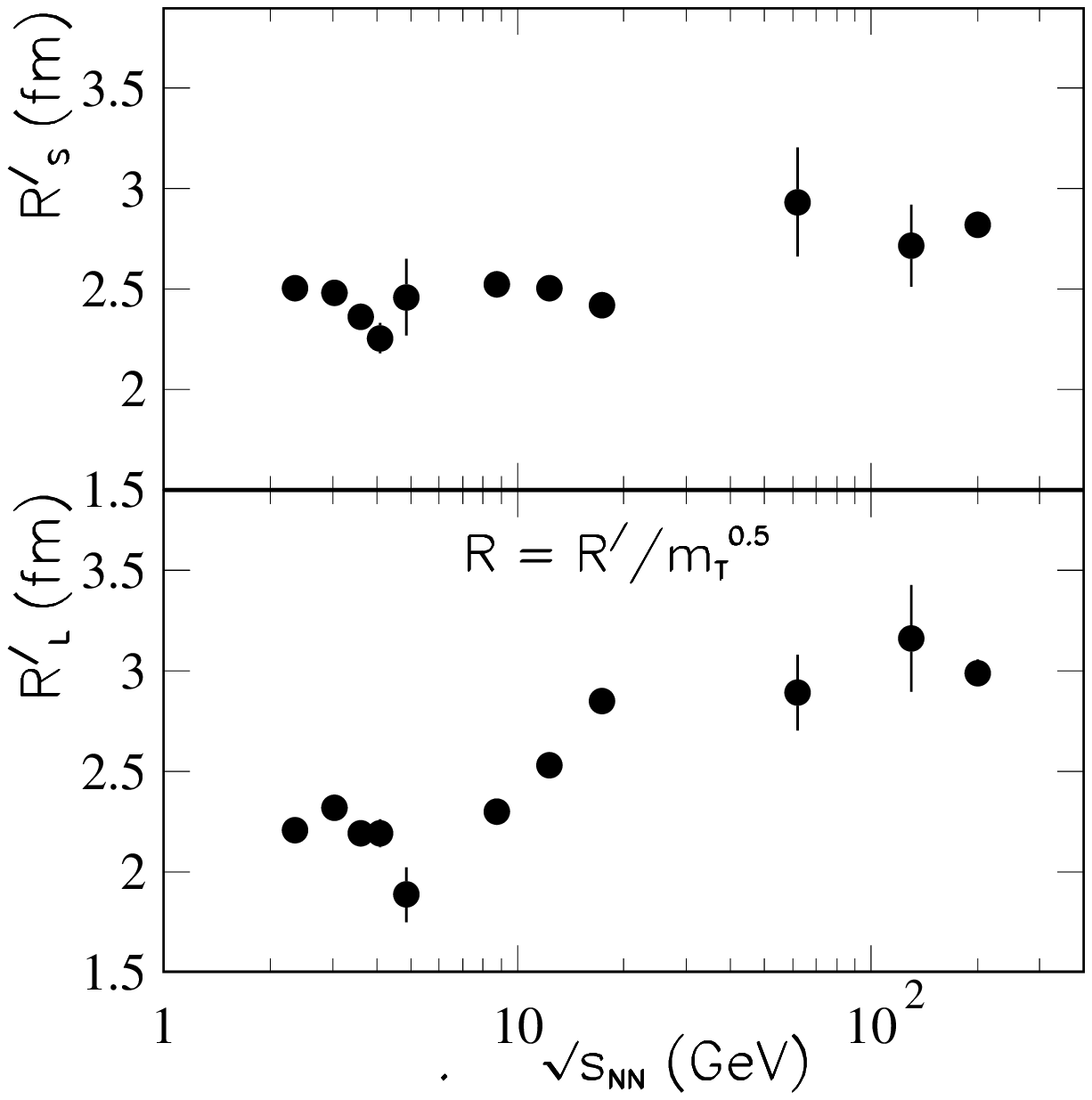}}
 \caption{\label{fig:RprimeFit}
 The excitation function of scale parameters $R^\prime_i$ from fits of the data 
 in Figure~\ref{fig:kTdepAllEnergies} to $R_i(m_T) = R^{\prime}_i/\sqrt{m_T}$.
 Note: this is a different functional form than that used in Figure~\ref{fig:PowerLawFits}.
 }
\end{minipage}
\end{figure}

Figure~\ref{fig:PowerLawFits} quantifies the evolution of the $m_T$-dependence
of the pion source radii with \roots,
using fits to $R_i(m_T) \sim m_T^{-\alpha_i}$.  As discussed in
Section~\ref{sec:TheoryLongitudinalFlow}, $\alpha_i = 0.5$ would represent
expectations for instantaneous thermal emission for a three-dimensionally
expanding fireball in the limit of large $m_T$.


The similarity persists as \Npart ~is varied.  In Au+Au collisions at RHIC, $R_i(m_T)$ ($i={\rm out, side, long}$)
simply scale as $N_{\rm part}^{1/3}$, with perhaps some flattening for $\Npart < 100$~\cite{Adams:2004yc,Adler:2004rq}.
Very similar $k_T$ dependence for different \Npart ~is also observed in Pb+Pb collisions at SPS~\cite{Adamova:2002wi} and
for Si+Au and Au+Au collisions at the AGS~\cite{Ahle:2002mi}.

In a flow-dominated freeze-out scenario,
the fall-off of transverse radii with $m_T$ increases as flow increases and/or temperature decreases~\citep[e.g.][]{Retiere:2003kf}.
Blast-wave fits to spectra~\cite{Xu:2001zj} indicate that freeze-out flow and temperature
vary significantly with \roots ~for $\roots \lesssim 10$~GeV.
The overall approximate \roots-independence of the $\alpha_i$
parameters may reflect the fact that significantly changing the slope of $R_i(m_T)$ requires very large
changes in flow and temperature; on the other hand, it could be that 
the compensating effects of smaller (larger) homogeneity lengths generated by larger flow (temperature) 
cancel almost exactly in nature.
Although $R_i(m_T)$ almost certainly reflects strong collective flow, determining
the strength of that flow requires other information, such as particle spectra~\cite{Lee:1990sk,Retiere:2003kf}.

because the radii fall off roughly as $1/\sqrt{m_T}$ (see Figure~\ref{fig:PowerLawFits}) and such a dependence
has been discussed frequently in the literature~\citep[e.g.][]{Beker:1995qv,Csorgo:1996no}, it is interesting to look at
the overall scale parameter from a single-parameter fit to  $R^\prime_i/\sqrt{m_T}$.  The \roots-dependence of
$R^\prime_{\rm side}$ and $R^\prime_{\rm long}$ are shown in Figure~\ref{fig:RprimeFit}.  The scale of the longitudinal 
homogeneity length grows significantly with \roots, consistent with an increase of the system evolution time.  However,
$R^\prime$ varies only very weakly with \roots.  

%


\subsubsection{Systematics with particle mass}
\label{sec:mTscaling}

Systematic studies for different mass particles provide additional controls
probing the space-time evolution of the source.  In particular for kaons, the
interpretation may be simplified owing to reduced effects of resonance
feed-down~\cite{Gyulassy:1989pp} and a reduced scattering cross-section for
$K^+$ in nuclear matter, raising the possibility that kaon correlations could
peer farther back to earlier stages of the
collision~\cite{Schnetzer:1982ga}.  Indeed, the first kaon
measurements~\cite{Akiba:1993cj,Beker:1994qs,Cianciolo:1995dd,Miskowiec:1996xa}
reported smaller source radii for kaons.  However, the observation that radii
for $K^+$ and $K^-$ were very similar~\cite{Beker:1994qs} was an early
experimental indication that different cross-sections were not the driving
physics behind these smaller radii.  This was supported by model
calculations~\cite{Sorge:1989vt}, which suggested that $K^+$ and $K^-$ in fact
scattered roughly equally in the dense medium created in heavy ion collisions.
In this case, the smaller radii for kaons results from their increased mass
in a flow field, not different cross-sections.

If indeed flow is generated in matter sufficiently dense that individual
cross-sections are unimportant, then all particles participate equally in
collective transverse flow.  In this case, their source radii should
approximately follow a common $m_T$
scaling~\cite{Akkelin:1995gh,Csorgo:1995bi,Wiedemann:1995au,Wiedemann:1999qn}.
Within uncertainties, first results on kaon interferometry by NA44 at the SPS
in S+Pb~\cite{Beker:1995qv} and Pb+PB~\cite{Bearden:2001sy} collisions were
consistent with a $1/\sqrt{m_T}$ scaling expected for isotropic Hubble
flow~\cite{Akkelin:1995gh,Csorgo:1995bi,Wiedemann:1995au}.  In some more recent
analyses~\cite{Afanasiev:2002fv}, the common $m_T$ systematic for the
transverse radii is less steep than for $\Rlong(m_T)$
(see Figure~\ref{fig:PowerLawFits}), as might be expected for more
boost-invariant (non-isotropic) flowing
systems~\cite{Wiedemann:1999qn,Retiere:2003kf}.

\begin{figure}[t!]
\centerline{\includegraphics[width=0.99\textwidth]{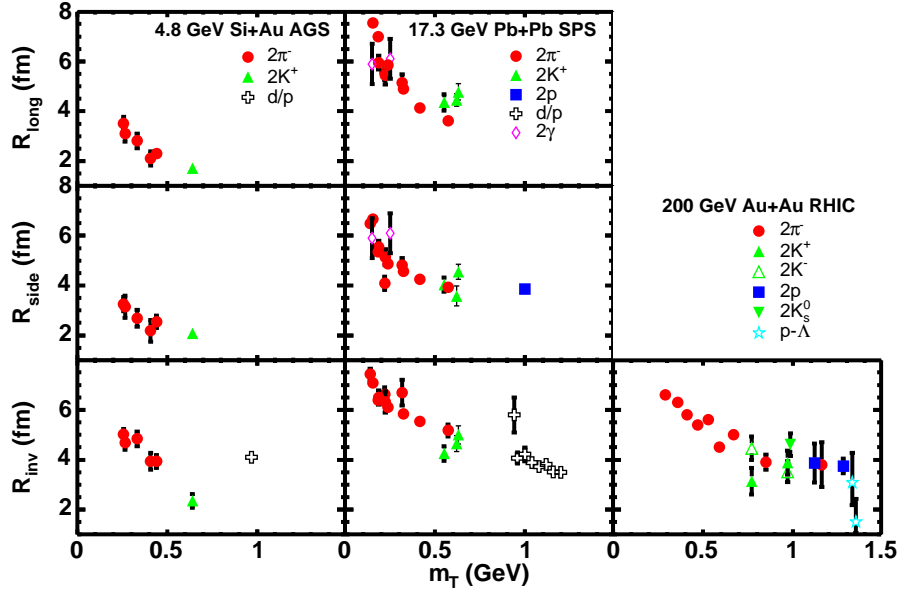}}
\caption{\label{fig:results_mt}
Transverse mass dependence of homogeneity lengths from correlations between particles of (nearly) identical mass.}
\end{figure}

Figure~\ref{fig:results_mt} collects the $m_T$ dependence of homogeneity lengths for several energies.
The left panels show results for Si+Au collisions at \roots=5.4~GeV, measured by E802 for
pions~\cite{Ahle:2002mi} and kaons~\cite{Akiba:1993cj,Cianciolo:1995dd}.
Femtoscopic radii for pions~\cite{Bearden:1998aq,Afanasiev:2002fv,Adamova:2002wi,Aggarwal:2002tm}
kaons~\cite{Bearden:2001sy}, protons~\cite{Appelshauser:1999in}, and photons~\cite{Aggarwal:2003zy}
measured in Pb+Pb collisions at the SPS are shown in the center panels.
The right panel shows the one-dimensional radius parameter $R_{\rm inv}$ measured at RHIC for pions, charged kaons,
and protons~\cite{Heffner:2004js}, neutral kaons~\cite{Bekele:2004ci}, and with
$\Lambda-p$ correlations~\cite{Renault:2004pp}.\footnote{Protons and $\Lambda$ baryons are not identical
particles, of course.  However, their masses are sufficiently close to try
including $R_{\rm inv}^{p-\Lambda}$ on a $m_T$-scaling plot.
Adherence to the scaling is consistent with a flow-dominated
scenario in which  homogeneity lengths depend only on mass.}
To compare across energies, $R_{\rm inv}$ results are included for the AGS and also for
the SPS, where the $R_{\rm inv}$ values were calculated from the 3D fit
results by accounting for the boost along the outwards direction from the LCMS (in which $P_z = 0$) to the
PCOM (pair center-of-mass, in which $\mid \stackrel\rightharpoonup P\mid = 0$) frame, $R_{\rm inv}^{2} = R_{\rm long}^2 + R_{\rm side}^2 + \gamma^2 R_{\rm out}^2$, where $\gamma$ is
given by $m_{T}/m$ of the pair.  Note that for massless particles, such as photons,
$\gamma$ is given by $k_{T}/Q_{\rm inv}$.  For a given orientation of the photon pair momentum 
$R_{\rm inv}$ can be related to 3D radii in the LCMS frame through the following expression for
the exponent of the Gaussian source term,
\begin{eqnarray}
\label{eq:gammaBoost}
Q^2_{\rm inv} R^2_{\rm inv} & = & Q^2_{\rm inv} \left( R^2_{\rm long}
	\cos^2\theta + R^2_{\rm side} \sin^2\theta sin^2\phi + k^2_T/Q^2_{\rm inv}
	R^2_{\rm out} \sin^2\theta cos^2\phi \right) \nonumber \\ 
& = & k^2_T R^2_{\rm out} \sin^2\theta cos^2\phi + Q^2_{\rm inv} \left( R^2_{\rm long}
        \cos^2\theta + R^2_{\rm side}
	\sin^2\theta sin^2\phi \right).
\end{eqnarray}
Thus \Rout~drops completely out of the source term and enters into the $\lambda$
coefficient~\cite{Aggarwal:2003zy}, but as a function of $k_{T}$.  For this reason
the 1D $Q_{\rm inv}$ fits from WA98 are plotted alongside values of \Rside~and \Rlong~instead.


This consistency between different particle types may carry an important message.
It calls into question theoretical scenarios which appear to explain $R_i(m_T)$ for particular particle
types only~\cite{Cramer:2004ih}.  
Further, the consistency with emission from a common flow-dominated source
may also support freeze-out scenarios in which the last scattering
in the dense phase determines the homogeneity region, instead of
milder rescatterings in the more dilute stage, which are dominated by
particle species-dependent cross-sections~\cite{Kapusta:2004ju,Wong:2004gm}. 

%
%
%
%
%
%

\subsection{New possibilities: Non-identical particle correlations}
\label{sec:systNonId}

Almost all femtoscopic measurements to date have been done through correlations of identical particles, usually charged pions.
With the availability of high-statistics data sets and new theoretical ideas~\cite{Lednicky:2001qv}, however,
experiments are beginning to make full multi-dimensional analyses of non-identical particle correlations.
These correlations are being used to test and refine the treatment of Coulomb effects in identical-particle correlation
analyses, to explore violations of flow-induced universal $m_T$ scaling, and to extract qualitatively new 
information on the space-time substructure of the source.

\subsubsection{Opposite-sign pion correlations}

The most common non-identical particle femtoscopy measurements have involved $\pi^+-\pi^-$ correlations.  Here, the natural
assumption is that the homogeneity regions of these particles coincide.  The primary interest is in whether the
homogeneity volume is extracted from $\pi^\pm-\pi^\pm$ correlations---driven by quantum statistics and Coulomb final state interactions (with minimal
strong FSI effects)-- is consistent with that extracted from $\pi^+-\pi^-$ correlations---determined only
 by Coulomb and strong
FSI~\cite{Adams:2004yc}. This issue has obvious implications for the FSI weight $F({\bf q})$ discussed in Section~\ref{sec:Fitting}.



At the SPS, $\pi^+-\pi^-$ correlations from reactions with sulphur~\cite{Alber:1997av} and lead~\cite{Appelshauser:1997rr} beams 
were consistent with emission from a homogeneity region with spatial scale roughly
the same as the radii from identical-particle
correlations.
Preliminary analysis of opposite-sign pion correlations from Au+Au collisions at the AGS~\cite{Miskowiec:1996xa,Miskowiec:1998ms}
similarly find consistency with scales extracted from like-sign pion correlations.
Recently, STAR reported~\cite{Adams:2004yc} strong consistency at RHIC energies as well.  Using large-statistics data sets,
they further show that contributions to the $\pi^+-\pi^-$ correlation function from strong-force interactions, though small,
are nevertheless important in explaining the data in detail.


\subsubsection{Other non-identical particle correlations in $|\vec{q}|$}
\label{sec:NonIdNoShift}

For statistical reasons, most non-identical particle correlations are measured
in one-dimensional $|\vec{q}|$-space, and thus probe only the average spatial
separation between the two particles at freeze-out in the pair
center-of-momentum frame (see Section~\ref{subsec:coulstrong}).  Because they
are sensitive to the size of each particles emission source and this separation
between them, it is often necessary to rely on the identical particle
correlations for a proper interpretation of results.


Despite these ambiguities, one-dimensional non-identical particle correlations may be used to
test existing systematics and expectations of the freeze-out scenario.
Conversely, if the freeze-out
geometry may be taken as given, non-identical particle correlations may place constraints on hadronic
scattering parameters, e.g. to measure the squared relative wave-function
of a pion and a $\Xi$ baryon.  First studies along these lines are underway at RHIC~\cite{Renault:2004me}.

Wang and Pratt~\cite{Wang:1999bf} suggested the measurement of $p-\Lambda$
correlations, which may be more sensitive to large structure than $p-p$
correlations, and which have a higher two-track reconstruction efficiency for some
experiments.  Published
results from Au+Au collisions at the AGS~\cite{Chung:2002vk} suggest a $p-\Lambda$ separation distribution with
a width similar to the proton homogeneity length and roughly consistent with $m_T$ scaling expectations
(cf Section~\ref{sec:mTscaling}).
Preliminary results at SPS~\cite{Blume:2002mr} and RHIC~\cite{Renault:2004me,Kisiel:2004it} give similar conclusions.
because the $p-\Lambda$ potential is not unambiguously known theoretically~\cite{Wang:1999bf}, it is unclear whether
possible statistically marginal violations of $m_T$ scaling~\cite{Chung:2002vk} or differences between baryon and
anti-baryon emission regions~\cite{Renault:2004me} are meaningful.  To first order, existing $p-\Lambda$ correlations
confirm existing systematics.

With the high-statistics and high-quality data sets at the largest collision energies,
truly exotic correlation studies are possible.  Preliminary
results~\cite{Kisiel:2004it} from RHIC on $\pi-\Xi$
correlations look particularly promising.  Here, the well-measured pion emission
distribution may be used to study the strange baryon freeze-out configuration.  It
may also provide information on the $\pi-\Xi$ final state interaction and scattering
cross-section which in turn can be used to constrain our understanding of the
sources of collective flow.  Blast-wave calculations reproduce the preliminary
$\pi-\Xi$ correlations, suggesting that the $\Xi$ flow is determined by its mass,
not its quark content.


%
%
%
\subsubsection{New information: non-identical particle correlations with directional cuts on $\vec{q}$}

Non-identical particle correlation analyses as a function of---or with cuts
on---the relative direction of $\vec{q}$ and $\vec{P}$ reveal qualitatively new
information~\citep[and 
  Section~\ref{sec:theoryBasics}]{Lednicky:1995vk,Voloshin:1997jh,Lednicky:2001qv}.
In particular, in addition to the root mean square (RMS) width of the separation distribution, the
direction and size of the average separation between the particles is
probed, although offsets in time and space cannot be disentangled; this is
shown as $\Delta r$ in Figure~\ref{fig:RvsMt}.  The correlation functions
selected for $\vec{q}\parallel\vec{P}$ and $\vec{q}\nparallel\vec{P}$ differ if
$\Delta r \neq 0$.  Furthermore, collective flow will induce position-momentum correlations detectable with directionally selected
non-identical particle correlations~\cite{Voloshin:1997jh,Retiere:2003kf}.

These correlations are statistically challenging, and few results are
available.  At RHIC, STAR has reported~\cite{Adams:2003qa} asymmetries in
$K^\pm-\pi^\pm$ correlations measured in central Au+Au collisions.  
Blast-wave calculations with transverse flow roughly adjusted to reproduce
other observations at RHIC~\cite{Retiere:2003kf} describe the data
semi-quantitatively.  Preliminary studies of $p-\pi$ correlations at the
SPS~\cite{Blume:2002mr} and RHIC~\cite{Kisiel:2004it} exhibit similar
mass-ordered spatial asymmetries in the transverse plane.  A preliminary study
at the AGS~\cite{Miskowiec:1998ms} reported very large ($\Delta r_{\rm
  long}\approx$~10~fm) average $p-\pi$ separations in the longitudinal
(beam) and impact parameter direction for forward-moving particles,
suggesting strong longitudinal flow; however, this result was never
confirmed.

We expect full three-dimensional analyses of a wide range of non-identical
particle combinations to be available in the near future from RHIC experiments.
Sophisticated analyses may probe non-trivial geometric substructure when
selecting on reaction plane, including the sideward shift predicted by
blast-wave calculations~\cite{Retiere:2003kf} when anisotropic flow structure
is present.

\section{Interpretations of Experimental Results}
\label{sec:interpretations}

In this section, we ask what we can learn from the spectrum of results just
presented.  Beginning with the broadest, least detailed observations, we move
to two fundamental quantities that may be directly extracted from the data,
and finally on to comparisons with specific models of heavy ion collisions.


\subsection{General Conclusions from Systematic Trends}

One of the first messages to take away from the discussion of
Section~\ref{sec:systematics} is that the results are stable across detector
and method; experimental systematics and uncertainties are under control.
Whatever difficulties we may have in interpreting measurements, we may be
confident that they do not have their origin in experimental artifact.

The size and shape inferred from two-particle correlations tracked with
collision geometry as anticipated.  Kinematic and mass dependences of
femtoscopic measurements showed the expected clear signatures of strong
collective flow in the beam direction and perpendicular to it.

\begin{SCfigure}[0.49][t!]
\caption{\label{fig:HBTexcitation}Panels (a-d): Excitation function of $\pi^-$
  source parameters at mid-rapidity and low $k_T$ ($\sim 0.17$~GeV/c) in central
  Au+Au(Pb+Pb) collisions.  PHENIX data are for $k_T \sim 0.26$~GeV/c and so
  fall somewhat below the trend.  Panel ({\it e}): Transverse freeze-out anisotropy
  parameter from non-central ($|\vec{b}|\sim 8$~fm) Au+Au collisions, estimated
  from the azimuthal dependence of source radii.  Data are
  from~\cite{Adcox:2002uc,Adler:2004rq, Adler:2001zd,
    Adams:2003ra,Adams:2004yc, Ahle:2002mi, Adamova:2002wi, Aggarwal:2002tm,
    Back:2004ug, Bearden:1998aq, Lisa:2000no,Lisa:2000xj}.  Also shown are
  calculations~\cite{Lisa:2000no,Lisa:2000xj,Bearden:1998aq,Hardtke:1999vf} at
  several energies with the RQMD model~\cite{Sorge:1995dp}; hashed region at
  other values of $\sqrt{s_{NN}}$ interpolates between these calculations.  }
        {\includegraphics[width=0.65\textwidth]{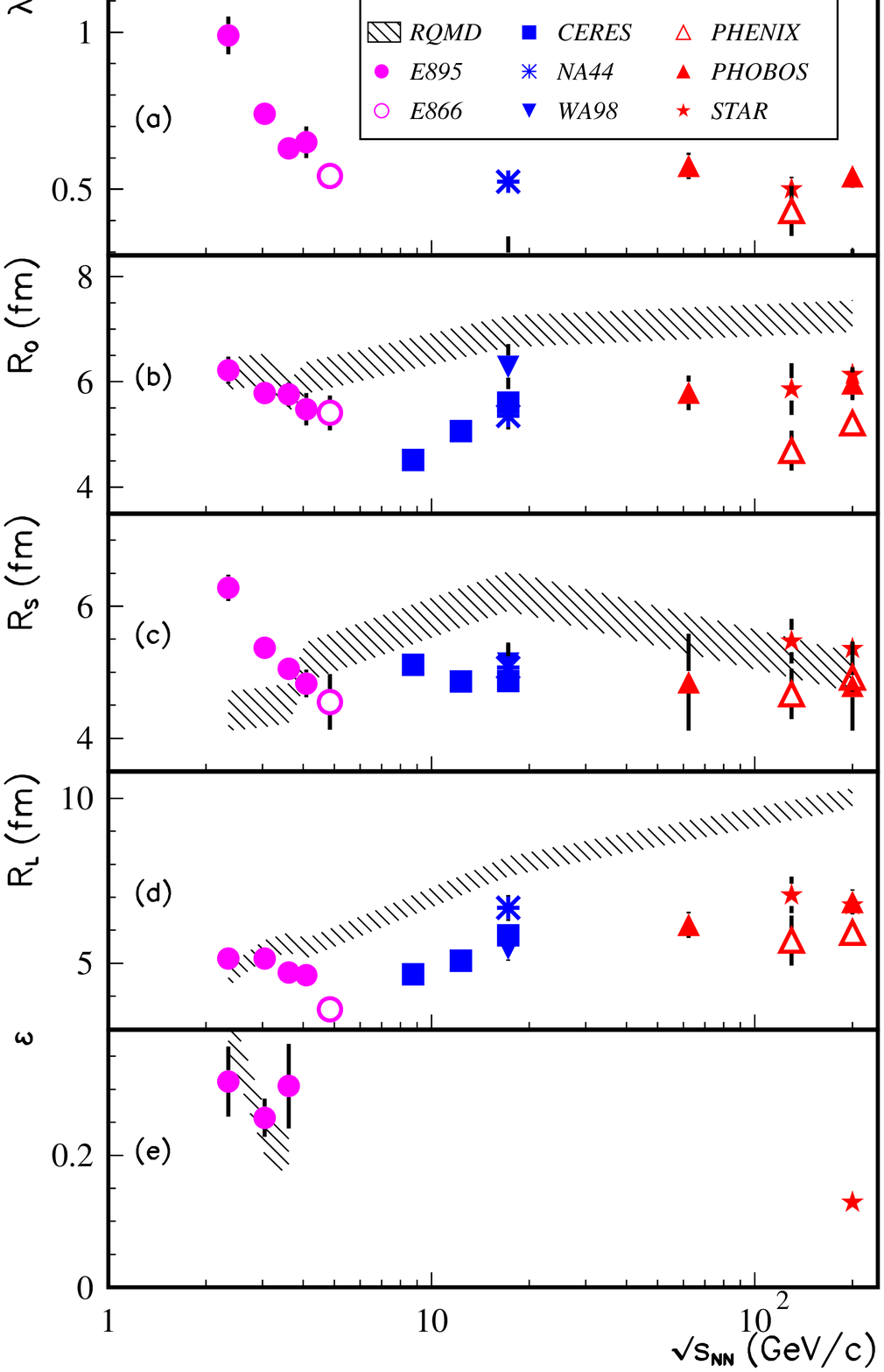}}
\end{SCfigure}

At a generic level (ignoring quantitative predictions), we are first given
pause at the jejune nature of the $\sqrt{s_{NN}}$-dependence of femtoscopic
parameters.  The most common example discussed is the pion source radii
excitation function shown in Figure~\ref{fig:HBTexcitation}.  In
Section~\ref{sec:systematics} we explored the trends in considerably
greater detail, but the figure conveys the right message: Scanning
$\sqrt{s_{NN}}$ through a range of two orders of magnitude changes final-state
geometry little.

Based on rather generic arguments of soft points in the equation of state or
entropy generation during a phase transition, there had been hopes for
non-trivial structure in the excitation function, as the energy threshold for
quark gluon plasma (QGP) creation was
crossed~\cite{Pratt:1986cc,Bertsch:1988db,Rischke:1996em,Harris:1996zx}.
General expectations were for long (relative to the explosion of a purely
hadronic system) system evolution time scales if QGP was formed. Blast-wave
analyses of data~\citep[e.g.][]{Retiere:2003kf} appear to rule out systems with
lifetimes in the neighborhood of 20~fm/c or higher.  However, owing to dynamic
effects, careful comparison with a dynamical model is required to extract
detailed evolution information; this is discussed in
Section~\ref{subsec:dynamicmodels}.

As discussed in Section 4.1, the weak increase with \Rout of the femtoscopic radii for central collisions shown in Figure 13 is mostly due simply to increasing multiplicity. The CERES collaboration~\cite{Adamova:2002ff}.              
has suggested that this increase may be understood in terms of an energy-independent mean free path of 1 fm at freeze-out. Interestingly, this explanation also appears to describe the decrease of radii with $\Rout~< 5 GeV$, when the different cross-sections and abundances of pions and protons are accounted for. This observation emphasizes the importance of chemistry in the determination of the freeze-out geometry. However, it neglects the dynamical structure of the source (e.g., flow) and the importance of the six-dimensional phase space density discussed in Section 2.9. We discuss this fundamental quantity next.

\subsection{Phase Space Density and Entropy}
\label{subsec:psdentropy_results}

\begin{SCfigure}[0.49][t!]
\includegraphics[width=0.5\textwidth]{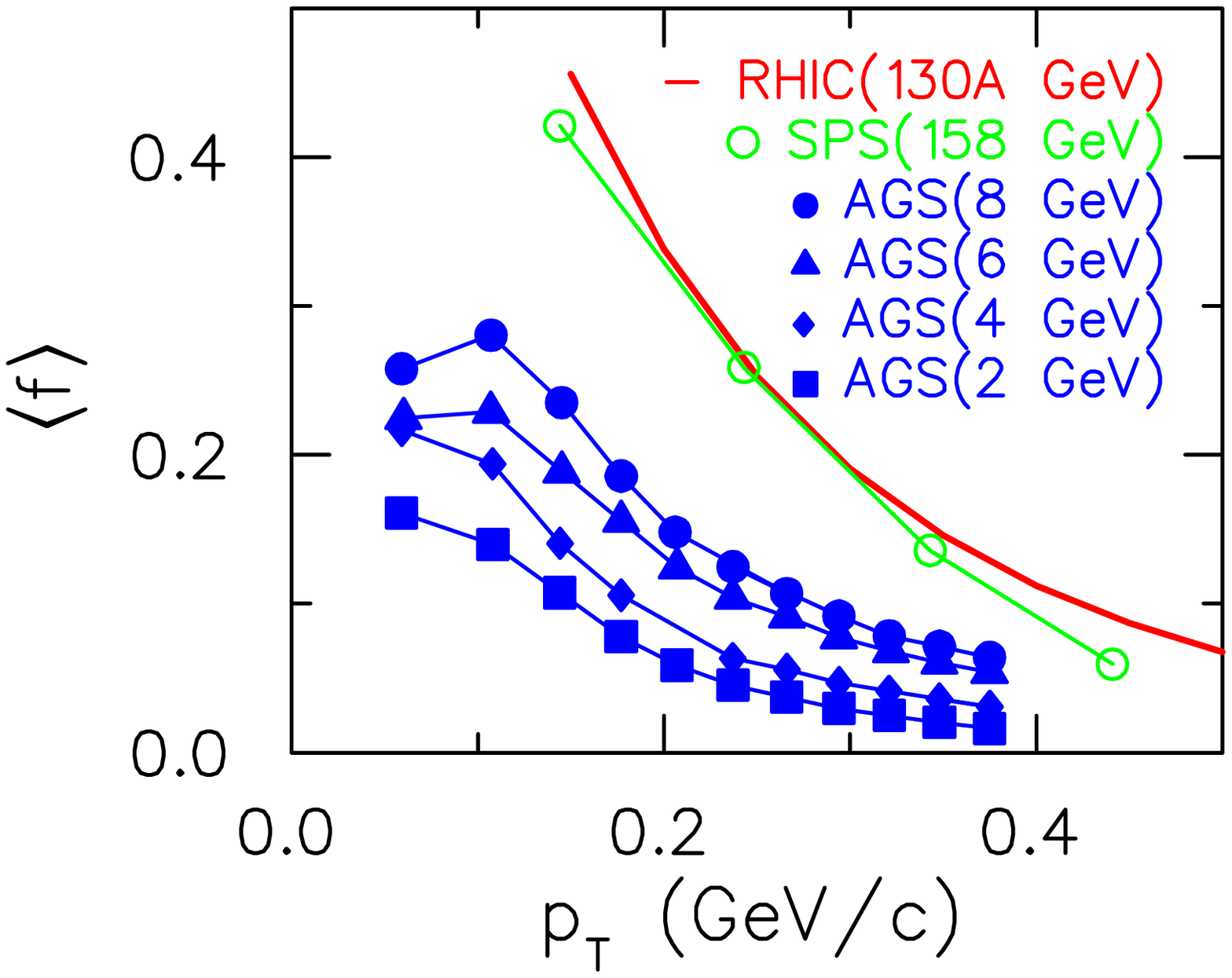}
\caption{\label{fig:psd} Average pionic phase space densities for central Au+Au
  and Pb+Pb collisions from the AGS to RHIC rise with beam energy but seem to
  plateau at SPS energies. Values were calculated with
  Equation~\ref{eq:fbargauss}.}
\end{SCfigure}

As shown in Section~\ref{subsec:psdentropy}, average phase space densities,
$\bar{f}({\bf p})$, can be calculated by combining source-size measurements
with spectra. Pionic phase space densities have been estimated for 130$A$ GeV
collisions at RHIC \cite{Pal:2003rz}, at the top SPS energies, and for several
AGS energies \cite{Lisa:2001fw}. Figure~\ref{fig:psd} shows results for all
three regions. In each case, the phase space densities were calculated via
Equation~\ref{eq:fbargauss}. For the SPS case, results were generated by applying
Equation~\ref{eq:fbargauss} to published spectra \cite{Afanasiev:2002mx} and
source-size measurements \cite{Kniege:2004pt}. We note that our calculations
for the SPS are higher than previously published values at low $p_t$
\cite{Ferenc:1999ku}. This discrepancy is likely due to the fact that in
Reference \cite{Ferenc:1999ku}, analytic parameterizations were used 
that differ significantly from published spectra at low $p_t$.

When applying Equation~\ref{eq:fbargauss}, an issue arises as to whether one should
subtract the contribution from resonances to the spectra. Indeed, if the pions
are created by decays so far outside the source volume that they do not
contribute to the correlation function, they should not be considered as pions
when calculating the phase space density. There are two strategies to correct
for such pions. First, one could use spectra where such particles are
subtracted and apply Equation~\ref{eq:fbargauss} literally. As a second option,
one could use the spectra without subtractions, but then multiply the
expression for $\bar{f}$ by $\sqrt{\lambda}$. because most published spectra have
been purged of the products of weak decays, the first method is usually
applied. However, published spectra still include the contribution from $\eta$'s,
which decay thousands of fm away from the source. The $\eta$ contribution was
accounted for in the SPS calculation for Figure~\ref{fig:psd} by reducing the
spectra by 5\%.

The phase-space densities in Figure~\ref{fig:psd} show a steady rise with beam
energy that seems to plateau at SPS energies. because the displayed phase space
densities have been averaged over coordinate space, the peak values are higher
\cite{Tomasik:2001uz,Tomasik:2002qt}, by a factor of $2\sqrt{2}$, if the spatial
profiles are Gaussian.  For a breakup temperature of 110 MeV, this requires a
rather high chemical potential, near or above 80 MeV
\cite{Greiner:1993jn,Pratt:1999ku,Akkelin:2004he}.

Ratios of particle yields are consistent with chemical
freeze-out at temperatures near 170 MeV
\cite{Braun-Munzinger:2003zd,Braun-Munzinger:2001ip,Andronic:2003zv}. This
suggests an interpretation where at higher densities the system is so strongly
interacting that yields equilibrate until the system reaches this temperature
and are then frozen during the subsequent freeze-out.  At AGS energies, the
increase in pion production brought on by increased beam energy results in more
pions being pushed into a given amount of phase space as realized in
Figure~\ref{fig:psd} by the rising phase space density. However, one would expect
that as the excitation energy surpasses the threshold for a 170 MeV
temperature, the phase space density would reach a limiting value. because the
average phase space density is preserved in an isentropic expansion with fixed
particle number (not exactly true if several masses are mixed together), one
would expect phase space densities to saturate once energy densities reached
this value. Indeed, the behavior in Figure~\ref{fig:psd} is consistent with this
scenario.

As shown in Section~\ref{subsec:psdentropy}, entropy can be calculated from
average phase space densities using Equation~\ref{eq:dsdy}. Spectra and
source-size measurements for baryons and mesons were used to estimate phase
space densities and entropy for 130$A$ GeV Au+Au collisions at RHIC. The total
entropy in the central unit of rapidity was estimated at $dS/dy=4450\pm 10\%$
\cite{Pal:2003rz}. In a hydrodynamic expansion, entropy is conserved though
viscosity, and shock waves might result in a roughly 10\% increase during the
evolution. Thus, this measurement provides an upper bound for the entropy at
$\tau\sim 1$ fm/c when thermalization first occurs. At $\tau=1$ fm/c, the
volume for particles in this rapidity slice is determined by the geometric
cross-sectional areas of the overlapping gold nuclei multiplied by $c\tau$,
thus an upper bound for the entropy density can be determined $s\le
(dS/dy)/(\tau\pi R^2)$. Knowing the energy density at $\tau=1$ fm/c would then
provide a point in the equation of state, $s$ vs. $\epsilon$. The value of
$dS/dy$ estimated in \cite{Pal:2003rz} is consistent with lattice calculations
if $\epsilon(\tau= 1~{\rm fm/c})\sim 7$ GeV/fm$^3$. Estimates of the original
energy density from the final-state measurement of $dE_t/d\eta$ are in the range of
4.5 GeV/fm$^3$ \cite{Adcox:2001ry}, but because these estimates neglect losses
from longitudinal work or the energy from longitudinal thermal motion, the 7
GeV/fm$^3$ value seems reasonable.


\subsection{Dynamic Models and their Comparison with Data}
\label{subsec:dynamicmodels}

It is increasingly recognized that the comparison of dynamic models of heavy
ion collisions to data is only insightful if it involves a sufficiently large
variety of experimental data. A comparison of dynamic models to femtoscopic
measurements alone (or to any other class of measurements alone) is of limited
value, simply because for realistic models, the number of possible
model-dependent parameter choices then tends to exceed the number of
experimental constraints.  In fact, all the model results that we review in
the current subsection remain unsatisfactory with this respect: They either
deviate significantly from femtoscopic data, or they reproduce these data at
the price of missing other important experimental information. In particular,
there is so far no dynamically consistent model that reproduces quantitatively
both the systematic trends discussed in Section 4 and the corresponding single
inclusive spectra. In this situation, the scope of this subsection is somewhat
limited. We want to explain why a dynamical understanding of femtoscopic
measurements is important. We shall also discuss the key physics input that
enters current attempts of dynamic modeling and the uncertainties resulting
from it. However, we shall try to bypass as far as possible model-dependent
details and rather focus on the question of which qualitative changes in the
underlying dynamics result in characteristic changes of femtoscopic data.

Correlation measurements provide a snapshot of the geometrical
distribution of particles at the time they decouple from the reaction. This
geometrical distribution provides a unique test of the dynamical evolution of
the produced matter at the late stage. because the spatial extension, dynamical
evolution, and lifetime of the produced system determines phase space density
and thus particle reaction rates, any dynamic model for the
latter has to be consistent with femtoscopic information. So far, correlation
analyses have focused mainly on Boltzmann (or cascade) models, on hydrodynamic
models, or on combinations of both (hybrid models). These model classes
correspond to rather different equations of state.

\begin{figure}[t!]
\centerline{\includegraphics[width=0.9\textwidth]{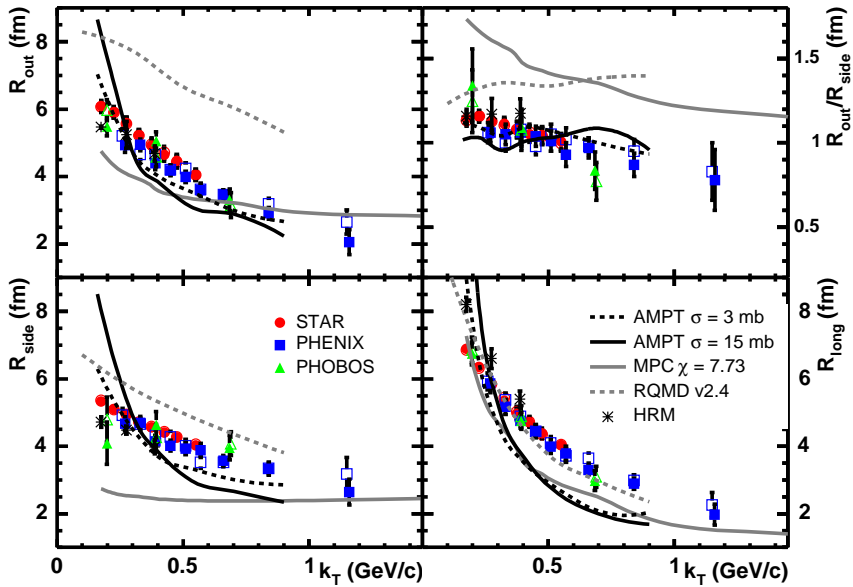}}
\caption{\label{fig:models_transport} Pion radius parameters from four Boltzmann/cascade models are compared to experimental RHIC data. For Molnar's Parton Cascase calculations, the reported radii are simply space-time variances. For the other models, radii are obtained from Gaussian fits to correlation functions generated according to the methods described in Section 2.8. Measured data for positive and negative pions are indicated by closed and open symbols, respectively.}
\end{figure}

The equations of state represented by cascade models tend to be stiff unless
they incorporate large number of resonant scatterings. If particles collide via
$2\rightarrow 1\rightarrow 2$ processes where the intermediate state has a
finite lifetime, the equation of state can be softened
\cite{Danielewicz:1995ay,Larionov:2001zg}. A prominent example of a cascade
model is RQMD \cite{Sorge:1989vt}, Relativistic Quantum Molecular Dynamics; it
is the only one which has been compared to data at AGS, SPS, and RHIC.  Results
for RQMD are shown in Figure~\ref{fig:HBTexcitation}.  However, for RHIC
energies, the dynamic consistency of RQMD is questionable because for most of its evolution, the model uses
 hadronic degrees of freedom, although RQMD simulations
for RHIC yield an energy density which stays above that of normal nuclear
matter for a significant duration ($\sim 5$ fm/c).  As can be seen in
Figure~\ref{fig:models_transport}, RQMD, which models the expansion as a
hadronic cascade, overpredicts $R_{\rm out}$ and $R_{\rm long}$ at RHIC
despite the fact that it underpredicts multiplicities. Another hadron cascade
model, the Hadronic Rescattering Model
\cite{Humanic:2002iw,Humanic:2003gs}, based solely on hadronic rescattering
with sudden collisions, gives smaller sources, which
come closer to the data. Part of the difference between the two hadronic
cascades may derive from RQMD's treatment of scatterings as resonant
interactions with finite lifetimes, which differs from the instantaneous
collisions employed in the Hadronic Scattering
Model. Figure~\ref{fig:models_transport} also displays results for Molnar's Parton Cascase (MPC)
\cite{Molnar:2002bz}, which aims to provide a transparent partonic toy model
by modeling the collision of light partons which undergo a one-to-one
hadronization to pions.  MPC, which has only instantaneous two-to-two
scatterings, should have the stiffest effective equation of state and
underestimates the source radii. 

On of the most complicated transport codes available is A Multi-Phase Transport (AMPT) model. AMPT \cite{Lin:2002gc,Lin:2003iq} aims at a
realistic description of all aspects of the reaction dynamics and includes a
partonic cascade coupled to a hadronic cascade employing a full list of
resonant interactions. AMPT provides a good fit to experimental radii, though
the $k_T$ fall-off is stronger than that found in the data, which fall off
$\sim m_T^{-1/2}$. The more rapid decrease of radius parameters with respect to
$k_T$ may be due to the continuous surface-like emission characteristic of
microscopic models, which when combined with cooling gives a higher relative
weight for high-energy particles to have been emitted earlier in time, before
the reaction volume has reached its full spatial extent. Any modification that
would reduce this type of emission should improve agreement with
data. Cautiously, one would conclude that the results in
Figure~\ref{fig:models_transport} favor models with a stiff, but not too stiff,
equation of state and no latent heat.

\begin{figure}[t!]
\centerline{\includegraphics[width=0.9\textwidth]{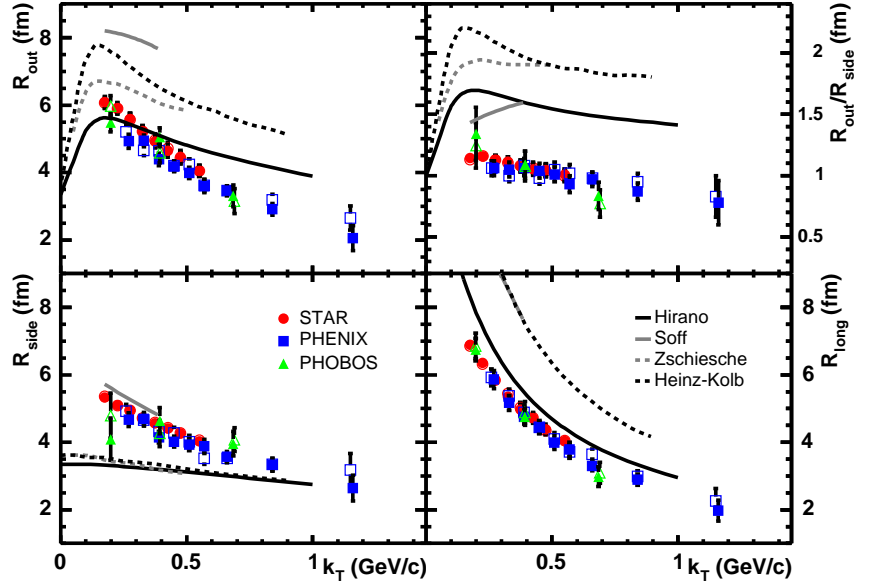}}
\caption{\label{fig:models_hydro} Hydrodynamic (Zschiesche, Hirano, and Kolb)
  and hybrid hydrodynamic/cascade (Soff) models calculations in comparison to
  RHIC data. Data are for 2$\pi^-$(open symbols) and 2$\pi^+$(closed symbols)
  source radii.}
\end{figure}

This conclusion is further supported if one compares the results of cascade
models with those of hydrodynamic simulations or hybrid models.
Figure~\ref{fig:models_hydro} compares results from RHIC to a three-dimensional
hydrodynamic model of Hirano~et al.~\cite{Hirano:2002ds} that
investigates the effects of resonance decays on chemical composition, the
two-dimensional model of Heinz and Kolb~\cite{Heinz:2002un}, and a two-dimensional
chiral model by Zseische~et al.~\cite{Zschiesche:2001dx} that performs
calculations for both first order and cross-over transitions.  In all cases, the
more favorable calculations are compared with the data: partial chemical
freeze-out for Hirano, and cross-over transition for Zschiesche, but large
discrepancies between the models and the data still remain. The results from Zseische were taken from the cross-over transition for the lowest critical temperature, 80 MeV. Substituting a first order phase transition or increasing the critical temperature to 130 MeV increased the value of $R_{\rm out}$ relative to $R_{\rm side}$, thereby increasing the differences with the data. Hirano also reported a calculation under complete chemical equilibrium at freeze-out; this improves the agreement with v$_2$, but at the expense of a much larger value of $R_{\rm out}$/$R_{\rm side}$.

In general, these models invoke equations of state that are
typically softer than those used in cascades and Boltzmann calculations, and
they often have latent heats to accompany the transition from the partonic
phase. As a consequence, lifetime and emission duration of the produced matter
are significantly larger than what one finds in cascade models, and such models
often significantly over-predict $R_{\rm long}$ and $R_{\rm out}/R_{\rm
  side}$. The fact that $R_{\rm side}$ comes out smaller than the data in
Figure \ref{fig:models_hydro} is mainly due to an attempt to compensate within
the available model-parameter range the very large time-scales as much as
possible. Purely hydrodynamic models can vary overall source volumes by
adjusting the break-up criteria, but doing so can make it difficult to fit the
three source dimensions, and their $m_T$ dependence, as well as other
observables. Several prescriptions have been applied to improve the modeling of
the breakup in the late stage to depend on microscopic considerations
determined by free-space cross-sections without having an extra adjustable
parameter, e.g. break-up density,
\cite{Csernai:2004pr,Sinyukov:2002if,Tomasik:2002qt}. An alternative to improve
the description of breakup is the use of hybrid models, in which hydrodynamic
evolution in the early stage is combined with cascading in the late stage.  In
Figure~\ref{fig:models_hydro}, we compare results of one such hybrid model, Ultrarelativistic Quantum Molecular Dynamics (URQMD)~\cite{Soff:2000eh,Bass:2000ib}, to results of hydrodynamic calculations.
Similar results were obtained by Teaney et al.
\cite{Teaney:2001gc,Teaney:2001av}. Compared to hydrodynamic simulations,
hybrid descriptions do not seem to notably reduce the overpredicted lifetimes.
They tend to emit most of the pions at a time near or above 15 fm/c and
significantly overpredict $R_{\rm out}/R_{\rm side}$ ratios, whereas blast-wave
parameterizations favor breakup times near or slightly below 10 fm/c.  However,
firm conclusions that the relative failure of hybrid models derives from the
chosen equations of state cannot be made until comparisons are made between
Boltzmann and hybrid calculations that use the same equation of state. Until
such an analysis is performed, other issues will cloud the interpretation, such
as whether viscous effects or details of the hydrodynamic/Boltzmann interface
dominate results and might even invalidate the hydrodynamic approach.

Entropy and pressure are intimately related, in that knowing the entropy
density as a function of the energy density determines the pressure as a
function of energy density. Whereas one can intuitively understand why a lower
pressure would lead to longer lifetimes and larger source dimensions, the
manifestations of changing the entropy are less transparent. One way to
understand the effects of changing the entropy is to associate higher entropy
with a larger effective number of light degrees of freedom. For instance, if
one were to hadronize via a one-to-one parton-to-pion scheme, the volume per
identical particle would change by a factor of the number of degrees of
freedom. If the r.m.s. momentum is the same before and after hadronization, this
would imply a change in entropy per particle equal to the logarithm of the
ratio of effective degrees of freedom, and if $\sim40$ light partonic degrees
of freedom were immediately replaced by three pionic degrees of freedom, the
system would lose $\ln(40/3)$ units of entropy per particle. To conserve
entropy, a system must expand its volume by a similar ratio which implies an
increase in radius parameters. Hydrodynamic models, which manifestly conserve
entropy, use the energy stored in the latent heat to provide the heat necessary
to preserve entropy during hadronization. Entropy conservation is more
difficult to enforce in microscopic approaches. This underlines the challenges
involved in applying microscopic simulations in an environment of strongly
interacting matter with ill-defined degrees of freedom, and it re-emphasizes the
importance of gaining a better understanding of the validity of hydrodynamic
calculations in such manifestly finite situations.  It is peculiar that the
entropy extracted from source-size and spectral measurements is consistent with
the lattice-inspired equation of state \cite{Teaney:2001gc,Teaney:2001av},
whereas the source sizes extracted from hybrid models incorporating similar
equations of state significantly overpredict source sizes. Part of this
contradiction can be explained by increases in the populations of baryons,
which, because they inherently have more entropy per particle than do pions, can
account for much of the missing entropy \cite{Pal:2003rz}. Thus, the HBT puzzle
does not necessarily imply an entropy puzzle.


The HBT puzzle is not so much that radius parameters cannot be fit by models,
but that our most sophisticated models, which incorporate a phase transition,
fail to reproduce the data. The very gradual evolution of extracted source
sizes as beam energies traverse a large range of energies is remarkable and
puzzling in its own right. A simple explanation is that the equations of state
do not dramatically change as the energy density changes from hadronic to
super-hadronic densities, i.e., there is not even a hint of a latent heat. In fact, the failure of hydrodynamical models to reproduce femtoscopic measurements might be largely due to unrealistically large latent heats needed to reproduce elliptic flow data~\cite{Pasi}.
However, as emphasized above, a host of unresolved issues prevent more
quantitative conclusions. These qualifiers cannot be lifted
until much more thorough analyses of models are performed that entail a
systematic exploration of the sensitivity of model predictions to both
parameters and assumptions. This would necessitate a tremendous commitment from
the community, but without it, many conclusions about the matter created in
relativistic heavy ion collisions will remain vague.



\section{Summary}
\label{sec:summary}

Twenty-five years ago, the goal of femtoscopy was to demonstrate that one
could measure a hadronic length scale with correlations, and if a result was on
the order of a few Fermi, the analysis was deemed a success. In contrast, with
the improved accuracy of measurements, the enormous increase in statistics,
and the simultaneous development of phenomenology and theory, femtoscopy is now
considered a precision measurement. Ten percent deviations between theory and
experiment are now taken seriously as evidence that the spatio-temporal
description of a model is significantly flawed. At RHIC energies, all six
dimensions of the correlation function have been exploited to provide truly
three-dimensional insight into the phase space cloud for particles of any
momentum with any direction.

For relativistic heavy ion collisions, there were expectations that a transition
from hadronic to partonic matter might be accompanied by a large latent heat,
which would bring about a dramatic change in the dynamics as beam energies
traversed the range for exploring the mixed phase. Within this range, it was
expected that the latent heat and the associated softening of the equation of
state would manifest itself by slowing the explosion with lifetimes approaching
or exceeding 20 fm/c. The signal of the phase transition would have been an
increase of the effective lifetime for a range of beam energies followed by a
return to more explosive and shorter-lived reactions at even higher energies.

Extended lifetimes were not observed. Increasing beam energies from AGS to RHIC
indeed causes larger energy densities and higher multiplicities, which push
toward increasing the source volumes. However, much of this increase in
multiplicities is absorbed by higher phase space densities rather than larger
source sizes. Combined with the increasing strength of radial flow, which
provides smaller regions of homogeneity relative to the overall source volume,
the result is that the effective dimensions change remarkably little over a
wide range of beam energies. Furthermore, it appears that lifetimes of the
reaction never leave the neighborhood of 10 fm/c. Not only does this represent
a lack of evidence for a latent heat, it also suggests that there is no
such latent heat.

These conclusions remain only modestly guarded. Theory and phenomenology are
progressing, but improvements in such aspects as mean-field effects or
accounting for the smoothness approximation are not expected to change
conclusions by more than 10\%. As with the improvements in including Coulomb
effects in Section \ref{sec:systematics}, removing some of the distortions and
aberrations from the analyses is likely to be significant for fine-tuning
models, but should not alter the conclusion that there is no large latent heat
associated with the reaction.

The theory of modeling, i.e., generating the source functions, has the greatest need for progress. because femtoscopic measurements are
determined solely by the geometry of breakup, changes in chemical or kinetic
evolution may have a significant impact on extracted source dimensions. The
next generation of transport theories should be more flexible and will
probably incorporate numerous effects such as in-medium mass changes, in-medium
reduction of scattering cross-sections, viscous effects, and dynamical solutions
for chemical rates.

Despite the progress listed above, measurement has only begun to address the
rich expanse of information available in correlations. Nearly all the
three-dimensional analyses have been focused on identical-pion
correlations. The huge data sets of the recent and upcoming runs at RHIC make
it possible to analyze source functions for many pairs of particles in six full
dimensions. In addition to providing important verification of identical-pion
measurements, these analyses address other issues, such as whether all species
flow and break up together.

This is not a field for the complacent. As emphasized above, efforts at RHIC
are just beginning to explore wholly new classes of correlations. Energy
densities at the LHC might surpass those at RHIC by the same factor that those
at RHIC surpassed the AGS. Just as our visions of the future from
twenty years ago proved largely naive, we should be prepared to be surprised
with the femtoscopy of the next 25 years.

\section*{Acknowledgements}



This work was supported by U.S. Department of Energy, Grants
DE-FG02-03ER41259 and W-7405-ENG-48, and by U.S. National Science Foundation
Grant PHY-0355007. The authors thank David Brown, Ulrich Heinz,
Sergei Panitkin, Nu Xu, and Kacper Zalewski for sharing insights.
For providing and explaining data, we thank
Harry Appelsh\"auser,  Giuseppe Bruno, Mike Heffner, 
Mercedes L\'{o}pez-Noriega, Michael Murray
and the experimental collaborations who conscientiously post data in numerical
form and respond to querries.
We thank Mark Meamber of the Physics Art
Team at Lawrence Livermore National Laboratory for creating
the illustrative figures.

\begin{twocolumn}


\end{twocolumn}

\end{singlespace}

\end{document}